\documentclass[12pt]{iopart}

\usepackage{iopams}
\usepackage{graphicx}
\begin{document}

\title[First passage leapovers and absorbing boundary conditions]{First passage leapovers of L\'evy flights and the proper formulation of absorbing boundary conditions}

\author{Asem Wardak}

\address{School of Physics, University of Sydney, New South Wales 2006, Australia}
\ead{mwar3732@uni.sydney.edu.au}
\vspace{10pt}
\begin{indented}
\item[]\today
\end{indented}

\begin{abstract}
An important open problem in the theory of L\'evy flights concerns the analytically tractable formulation of absorbing boundary conditions.
Although numerical studies using the correctly defined nonlocal approach have yielded substantial insights regarding the statistics of first passage, the resultant modifications to the dynamical equations hinder the detailed analysis possible in the absence of these conditions.
In this study it is demonstrated that using the first-hit distribution, related to the first passage leapover, as the absorbing sink preserves the tractability of the dynamical equations for a particle undergoing L\'evy flight.
In particular, knowledge of the first-hit distribution is sufficient to fully determine the first passage time and position density of the particle, without requiring integral truncation or numerical simulations.
In addition, we report on the first-hit and leapover properties of first passages and arrivals for L\'evy flights of arbitrary skew parameter, and extend these results to L\'evy flights in a certain ubiquitous class of potentials satisfying an integral condition.
\end{abstract}

\vspace{2pc}
\noindent{\it Keywords}: L\'evy flights, first passage, leapovers, stochastic analysis

%\submitto{\jpa}

%\maketitle

\section{Introduction}

% par1:
% - what are levy flights (metzler)?
A growing body of work has uncovered the ubiquity of anomalous diffusion in natural phenomena \cite{Metzler_2004}.
Characterised by long-range correlations in space and time, anomalous diffusive processes distinguish themselves from classical Brownian diffusion by a nonlinear time dependence of the mean squared displacement $\langle x^2(t)\rangle\propto K_\gamma t^\gamma$, where $\gamma\neq1$, caused by the breakdown of the central limit theorem \cite{METZLER20001}.
In the continuous time random walk description, where the length of a jump along with the waiting time between two consecutive jumps is jointly described by a probability distribution, the case of a finite mean waiting time and an infinite jump length variance in the continuous long-term limit corresponds to L\'evy flight, an example of Markovian (memoryless) superdiffusion ($\gamma>1$).

% - what is the FP problem (blumenthal,getoor,rogozin,sokolov,lachal,kyprianou,dybiec2016,profeta,)?
L\'evy flights in the absence of boundaries are succinctly described in the presence of an arbitrary external potential by the fractional Fokker-Planck equation (FFPE) \cite{Klages2008}.
Also known as the Kolmogorov forward equation or the Smoluchowski equation, the FFPE is a fractional partial differential equation on the position probability density.
However, many physical, chemical, biological, astronomical, geological and meterological situations \cite{Metzler_2004,METZLER20001,Klages2008} are concerned with the passage of these dynamical processes across a boundary.
Unlike classical Brownian motion, the problem of first passage for L\'evy flights encounters a number of intricacies due to the discontinuous nature of its sample paths, or trajectories, leading to leapovers over the boundary at the time of first passage.
For processes in the half-line, the distribution of first hits is equivalent to the first passage leapover density (FPLD).
Early progress was made in obtaining the distribution of first hits, also known as the harmonic measure, out of both finite and infinite regions in the 1960s \cite{10.2307/1993561,10.2307/1993412} and 1970s \cite{doi:10.1137/1117035}, which was recently revisited in the mathematical \cite{kyprianou2014,Profeta2016} and physical \cite{Dybiec_2016} context.
So far, however, FPLDs of L\'evy flights have only been investigated for the symmetric and one-sided cases in the absence of a potential well.
Other Markovian pseudoprocesses with discontinuous paths have also been shown to display nontrivial first passage properties, with multipoles occuring in their first hit distributions \cite{lachal2007,LACHAL20081}.
That this causes issues for the derivation of first passage times is ultimately unsurprising, as detailed knowledge of the trajectory of the underlying random process is required to uniquely determine the first passage time density (FPTD) \cite{Sokolov_2004}, explaining the failure of the method of images which had previously succeeded in describing the first passage properties of Brownian motion \cite{0305-4470-36-41-L01}.

% par2: previous ways of accounting for BCs:
In order to determine the properties of first passage for L\'evy flights, a number of techniques have hitherto been used in order to account for the required absorbing boundary conditions.
% - method of images (wrong),
% - simulations (cop out) (zumofen,dybiec2017,dybiec2006),
Numerical simulations of L\'evy flights using the Langevin description \cite{PhysRevE.51.2805,PhysRevE.73.046104,PhysRevE.95.052102} can easily account for the problem of first passage by physically removing the particle when it first escapes the allowed region.
% - discretisation (cop out) (zoia),
By spatially discretising the fractional Laplacian operator (the infinitesimal generator of the L\'evy process driving the particle) using a matrix representation, the absorbing boundary condition was implemented by truncating the matrix outside the allowed region \cite{PhysRevE.76.021116}.
In this discretised case, the moments of the FPTD were recursively expressed in a manner that allowed for numerical approximation, with the accuracy determined by the size of the discretisation step.
Although discretisation and simulations provide a rough picture of the properties of first passage, they cannot replace a full analytical treatment where the first passage time is fully described along with the position density of the particle, as in the classical Brownian case.
% - truncation of the integral (difficult) (buldyrev,chechkin,deng),
It has been known for some time that the analytical formulation of the absorbing boundary condition in the FFPE is consistently phrased in terms of the truncation of the integral form of the fractional Laplacian operator \cite{0305-4470-36-41-L01}.
Using the Sonin inversion formula, this technique yielded an analytical form for average properties of symmetric L\'evy flights such as the mean first passage time (MFPT) \cite{PhysRevE.64.041108,BULDYREV2001148}.
Due to the lack of a convenient representation of this nonlocal integral operator in Fourier space, its computational intractability prevented this formulation from being used to determine the position density of the particle \cite{0305-4470-36-41-L01} or any details about the FPTD on the half-line apart from the universal Sparre-Andersen scaling of its tail \cite{Andersen_1953,ANDERSEN1954}.
However, it is still considered the standard formulation of the absorbing boundary condition in the FFPE \cite{doi:10.1137/17M1116222,PhysRevE.98.022208}.
% - asymptotics/joint FPLD-FPTD (eliazar,koren),
By using a description of the joint FPTD-FPLD in Laplace space \cite{ELIAZAR2004219} the first passage of one-sided L\'evy motions on the half-line were able to be analytically described.
Due to a theorem by Skorohod \cite{Skorokhod2004}, the prefactor of the tail of the FPTD for symmetric L\'evy flights in the half line was found along with its Sparre-Andersen scaling and the full FPLD \cite{PhysRevLett.99.160602,KOREN200710}.
More recently, these FPTD tail prefactor results were extended to L\'evy flights of arbitrary skew parameter on the half-line using the same theorem \cite{Padash_2019}.
% - perturbation scheme (zoia,garcia-garcia),
Perturbation schemes, inspired by quantum mechanics, have also been investigated around the Brownian solution in order to correctly predict the leading and subleading behaviour of the tail of the position density for symmetric L\'evy flights on the half-line \cite{PhysRevE.51.2805,PhysRevE.76.021116,PhysRevE.86.011101}.
Nevertheless, the correct and analytically tractable formulation of absorbing boundary conditions for general L\'evy flights in arbitrary regions, such that the FPTD and position density be analytically determined, remains an open problem.

% - first arrivals instead of FP (janakimaran, other search papers)
The difficulty of the problem of first passage for L\'evy flights has led to some groups turning to other variants of boundary problems for L\'evy flights.
The usage of a delta-function sink in the FFPE is known to describe the first arrival of the particle to a single point, regardless of the underlying trajectory of the random process; which differs from the first passage in all non-Gaussian L\'evy flights.
This case has been investigated in the context of search \cite{Palyulin2931}, and both the first arrival time density and position density have been obtained in Laplace space for L\'evy flights in free, linear and harmonic potential wells for arbitrary point sink strengths representing various probabilities of absorption \cite{PhysRevE.95.012154}.
Since first arrival is equivalent to first passage in the Gaussian case of Brownian motion, the same technique is used to formulate absorbing boundary conditions in classical Brownian motion.
A reverse kind of first arrival, stochastic resetting, has also been investigated \cite{evans2019stochastic}, where the particle is stochastically reset to a given region.
Finally, reflecting boundary conditions have had some discussion with regard to the implementation of its different realisations in L\'evy flights \cite{PhysRevE.95.052102}.
In particular, the motion stopping formulation was found to be equivalent to an infinite potential well with no other restrictions on the flight, leading to an easily solvable FFPE \cite{PhysRevE.77.061112}.
Overall, regardless of the type of boundary conditions imposed, the sources and sinks that have been employed in the dynamical equations have either been explicitly dependent on the position density itself, or of the single-point delta-function form.

% par3: first-hit distribution as the absorbing sink properly formulates the absorbing BC
In this paper, we show that using the first-hit distribution as the absorbing sink term in the FFPE correctly and efficiently formulates the absorbing boundary condition for general Markov processes in arbitrary regions, without modifying the key dynamical operators of the equation.
Given an arbitrary source $f(x,t)$ extended in space and time, we construct a canonical absorbing sink term $S_{x,t}[f(x,t)](x,t)$ which respects the absorbing boundary condition, and depends only on the FPTD and first-hit distribution of the process along with the source.
We obtain an equation for the MFPT of a general Markov process out of an arbitrary region using only the free propagator and the first-hit distribution.
In the case of free L\'evy flights of arbitrary skew parameter $-1\leq\beta\leq1$, we express the FPTD and position density of the particle in an exact fashion in Laplace space, dependent only on the free propagator and the first-hit distribution.
We unify the notions of first passage and first arrival by considering the arrival to a domain as equivalent to the passage out of its complement.
This allows us to use the same expressions to find first-arrival densities to arbitrary (not necessarily point-like) regions along with the corresponding position densities.
We apply the results to a number of interesting well-known test cases in order to demonstrate the power of this novel framework.
To do this, we review recent mathematical results on first hits for L\'evy flights of arbitrary skew and extend them to L\'evy flights on the half-line in external potentials termed \emph{boundary-centred potentials}.

% par4: the rest of the paper is organised as follows
The remainder of the paper is organised as follows.
In section \ref{sec:dybiec} we report on the first-hit properties of free L\'evy flights of arbitrary skew, focusing on escapes from and arrivals to intervals of half-width $L$.
In section \ref{sec:koren} we use these results to investigate the leapover properties of L\'evy flights on the half-line of arbitrary skew, in free and boundary-centred potentials, such as the harmonic potential well with the boundary at its local extremum.
In section \ref{sec:chechkin} we introduce the general framework outlined above.
We discuss the implications of this novel framework in section \ref{sec:discussion} and compare it in more detail to previous work in the area.

For clarity of expression, Fourier transforms $\langle e^{ikx}\rangle$ of spatial variables $x_{(\cdot)}$ are indicated by the explicit use of the variable $k_{(\cdot)}$, and Laplace transforms of temporal variables $t_{(\cdot)}$ are indicated by usage of the variable $s_{(\cdot)}$.
We also introduce a rescaled version of the skew parameter $\beta$,
\begin{equation}\label{eq:rescaledskew}
    \beta':=\frac{2}{\pi}\arctan\left(\beta\tan\frac{\pi\alpha}{2}\right)
\end{equation}
which is related to the positivity parameter \cite{Bingham1973,JeanBertoin2012} $\rho$ by the identity $2\alpha\rho=\alpha+\beta'$.

\section{First arrival and passage of free L\'evy flights in finite domains}\label{sec:dybiec}

In this section, we employ a number of results in the mathematical literature to explore spatial properties of one-dimensional L\'evy flights of arbitrary skew $-1\leq\beta\leq1$.
For easy comparison with previous studies in a physics context, we reproduce the situation set up in \cite{Dybiec_2016} for the symmetric case $\beta=0$.
This situation necessitates the rescaling of those mathematical results, the process of which is described in \ref{appdx:dybiec}.
We use these results to obtain the first-hit distribution (\ref{eq:firstescape_q}) corresponding to the first passage out of $[0,2L]$, which is equivalent to the first arrival to
$\mathbb{R}\setminus[0,2L]=(-\infty,0)\cup(2L,\infty)$;
along with the first-hit distribution (\ref{eq:firstarrival_q}) for the first arrival to $[-2L,0]$, equivalent to the first passage out of $(-\infty,-2L)\cup(0,\infty)$.

The formal setup is as follows \cite{10.2307/1993561,10.2307/1993412,doi:10.1137/1117035,kyprianou2014,Profeta2016}.
Let $X=\{X(t)~|~t\geq0\}$ be an $\alpha$-stable L\'evy process satisfying the stochastic differential equation $dX=dL(\alpha,\beta,\gamma,D)$, where $0<\alpha\leq2$ is the L\'evy stable index, $-1\leq\beta\leq1$ is the skew parameter, $\gamma\in\mathbb{R}$ is the centre, and $D>0$ is the scale parameter.
The classical Gaussian process is recovered when $\alpha=2$.
Since the skew parameter $\beta$ has a slightly different meaning when $\alpha=1$, we consider the case $\beta=0$ whenever $\alpha=1$ for the sake of simplicity, along with $\gamma=0$ for all $\alpha$.
We denote by $\mathbb{P}_{x_0}$ the law of $X$ starting from $x_0$, and the time of first passage out of the region $\Omega$ by $T_\Omega:=\inf\{t>0~|~X(t)\notin\Omega\}$.
The density of the harmonic measure $\mathbb{P}_{x_0}[X(T_\Omega)\in dx~|~T_\Omega<\infty]$, or equivalently the first-hit distribution, describes the relative probabilities of points visited at the time of first passage out of $\Omega$.
As a result, its support is contained in $\overline{\mathbb{R}\setminus\Omega}$.

\subsection{Escape from a finite interval and half-line}

We begin by considering a particle undergoing free L\'evy flight starting at the position $x_0\in[0,2L]=:\Omega$.
The distribution of first hits upon the escape of the particle out of the interval $[0,2L]$ has (normalised) density (see (\ref{eq:a1}))
\begin{equation}\label{eq:firstescape_q}
    q(x|x_0) = \frac{\sin(\pi\frac{\alpha+\mathrm{sgn}(x-L)\beta'}{2})}{\pi}
    \left|\frac{x_0}{x}\right|^\frac{\alpha-\beta'}{2}
    \left|\frac{2L-x_0}{2L-x}\right|^\frac{\alpha+\beta'}{2}
    \frac{1}{|x-x_0|}
\end{equation}
when $x\in(-\infty,0]\cup[2L,\infty)=\overline{\mathbb{R}\setminus\Omega}$, and $q(x|x_0)=0$ when $x\in(0,2L)=\mathrm{Int}(\Omega)$.
In the Gaussian limiting case $\alpha\to2$ the density $q(x|x_0)=(1-x_0/2L)\delta(x)+(x_0/2L)\delta(2L-x)$ is a sum of delta functions at either end of the allowed region, reflecting the continuous sample paths of the Gaussian process.
The weightings of these delta functions represent the probabilities that the particle will hit the respective ends of the boundary first, which in the Gaussian case $\alpha=2$ is linear in the initial position $x_0$.

This quantity can be used to characterise the first passage leapover length $l\geq0$, the distance between the crossed boundary and the first hit point, with normalised density $f(l|x_0) = q(-l|x_0)+q(2L+l|x_0)$.
Applying this to (\ref{eq:firstescape_q}) yields the FPLD
\begin{eqnarray}
    f(l|x_0) &= \frac{\sin(\pi\frac{\alpha-\beta'}{2})}{\pi}
    \left|\frac{x_0}{l}\right|^\frac{\alpha-\beta'}{2}
    \left|\frac{2L-x_0}{2L+l}\right|^\frac{\alpha+\beta'}{2}
    \frac{1}{|l+x_0|} \nonumber\\
    &\quad+ \frac{\sin(\pi\frac{\alpha+\beta'}{2})}{\pi}
    \left|\frac{x_0}{2L+l}\right|^\frac{\alpha-\beta'}{2}
    \left|\frac{2L-x_0}{l}\right|^\frac{\alpha+\beta'}{2}
    \frac{1}{|2L+l-x_0|}
\end{eqnarray}
when $l\geq0$, and $f(l|x_0)=0$ when $l<0$.
The FPLD decays as $l^{-(1+\alpha)}$ for all finite half-widths $L$, independent of the skew $\beta$.
In the half-line limit $L\to\infty$ (whereby the particle undergoes escape from $\Omega=[0,\infty)$), the notion of leapovers becomes equivalent to that of the first hit, and so
\begin{equation}\label{eq:fpl_dybiec}
    f(l|x_0) = q(-l|x_0) = \frac{\sin(\pi\frac{\alpha-\beta'}{2})}{\pi}
    \left|\frac{x_0}{l}\right|^\frac{\alpha-\beta'}{2}
    \frac{1}{|l+x_0|}
\end{equation}
where the FPLD in this limit decays as $l^{-(1+(\alpha-\beta')/2)}$, dependent on $\beta$.
This result is exhibited more generally as an interpolation in the asymptotics of the FPLD with respect to the half-width $L$:
\begin{equation}
    f(l|x_0) \propto \cases{
        l^{-(1+(\alpha-\beta')/2)}   &   for $x_0\ll l\ll L ~,$ \\
        l^{-(1+\alpha)}  &   for $x_0\ll L\ll l~.$ }
\end{equation}
As a result, the skew parameter $\beta$ is only important in the intermediate asymptotics of the FPLD, when $L$ is large compared to $l$.
This can be explained by the observation that when $L$ is small compared to the leapover length $l$, jumps of the process in either direction contribute indiscriminately to the FPLD, whereas in the case of large half-width $L$ relative to $l$, only those jumps of the process in the negative direction are directly responsible for the FPLD, breaking the directional symmetry and establishing the effect of the rescaled skew $\beta'$.

\subsection{Arrival to a finite interval and half-line}

We now investigate the first-hit properties of particles arriving at $[-2L,0]$ from outside the interval, which has a number of important differences from the previous case.
This corresponds to the first passage out of $\Omega:=\mathbb{R}\setminus[-2L,0]=(-\infty,-2L)\cup(0,\infty)$.
The first-hit distribution for such a particle starting at $x_0\in\Omega$ has density (see (\ref{eq:a2}))
\begin{eqnarray}\label{eq:firstarrival_q}\fl
    q(x|x_0)
    = \frac{\sin\left(\pi\frac{\alpha-\mathrm{sgn}(x_0+L)\beta'}{2}\right)}{\pi(2L+x)^\frac{\alpha+\beta'}{2}|x|^\frac{\alpha-\beta'}{2}} \Bigg(
    \frac{|2L+x_0|^\frac{\alpha+\beta'}{2}|x_0|^\frac{\alpha-\beta'}{2}}{|x-x_0|}
    \nonumber\\
    - \frac{\max(\alpha-1,0)}{L}\int_1^{|x_0+L|/L}(t-1)^{\frac{\alpha-\mathrm{sgn}(x_0+L)\beta'}{2}-1}
    (t+1)^{\frac{\alpha+\mathrm{sgn}(x_0+L)\beta'}{2}-1} dt
    \Bigg)
\end{eqnarray}
when $x\in[-2L,0]$, and $q(x|x_0)=0$ when $x\in\Omega$.
Because the $\alpha$-stable motion $X(t)$ is transient when $0<\alpha<1$, this density is only normalised when $1\leq\alpha\leq2$, and so the probability of not hitting the finite first arrival interval $[-2L,0]$ is nonzero when $0<\alpha<1$: (see (\ref{eq:a3}))
\begin{equation}
    R(x_0) = \frac{\Gamma\left(1-\frac{\alpha+\mathrm{sgn}(x_0+L)\beta'}{2}\right)}{\Gamma\left(\frac{\alpha-\mathrm{sgn}(x_0+L)\beta'}{2}\right)\Gamma(1-\alpha)}
    \int_0^\frac{|x_0+L|-L}{|x_0+L|+L}\frac{t^{\frac{\alpha-\mathrm{sgn}(x_0+L)\beta'}{2}-1}}{(1-t)^\alpha}dt ~.
\end{equation}
When $L\to\infty$, $R(x_0)\to0$, since in this case we recover the first arrival to the negative half-line, which is equivalent to the first passage out of the positive half-line examined in the previous subsection.
On the other hand, when $L\to0$, the classical problem of first arrival to a point is reconstructed and the point is almost certainly not hit by the particle: $R(x_0)\to1$.

The disconnectedness of the allowed region $\Omega$ has an important consequence for the computation of the finitely supported leapover density.
Since the particle can jump over the entire $[-2L,0]$ interval without landing inside it, the first-hit distribution does not provide sufficient information on which boundary was crossed between the first hit point and the particle's location immediately before the first hit point.
As a result, the FPLD cannot always be computed from the first-hit distribution alone.

\section{Leapover lengths for $\alpha$-stable processes with boundary-centred potentials on the half-line}\label{sec:koren}

Another scenario of interest in characterising the first-passage properties of L\'evy flights (e.g.\ \cite{PhysRevLett.99.160602,KOREN200710} in the symmetric case $\beta=0$) consists of free L\'evy flight out of the half-line.
By using the results reported in the previous section, it is possible to generalise the results on leapover lengths in this scenario to the case of arbitrary skew $-1\leq\beta\leq1$.
Remarkably, using a variable transformation, these results also yield FPLDs for general L\'evy flights on the half-line in a wide class of external potentials characterised by the boundary residing at a natural centre of the potential function, which we term \emph{boundary-centred potentials}.

\subsection{Leapover lengths for free $\alpha$-stable processes}

We first reproduce the setup from \cite{PhysRevLett.99.160602} for the symmetric case $\beta=0$ and extend it to the case of arbitrary skew $-1\leq\beta\leq1$.
Note that the extension of the FPTD in this scenario is covered in section \ref{sec:chechkin} and given by (\ref{eq:freelevyFPTD}).
A particle undergoing free L\'evy flight starting at the position $x_0=0$ escapes from the region $(-\infty,d)$, and we are interested in its FPLD.
Reflecting the space axis, we obtain the escape from $(-d,\infty)$, where $x_0=0$ (equivalent to the escape from $(0,\infty)$ where $x_0=d$ by translational symmetry of the free L\'evy process), allowing us to use the results from the previous section.
However, since the escape is in the opposite direction, the sign of the skew parameter $\beta$ must be flipped when adapting the expressions.
Adapting (\ref{eq:fpl_dybiec}) we obtain
\begin{equation}\label{eq:fplkoren}
    p_d(l) = \frac{\sin(\pi\frac{\alpha+\beta'}{2})}{\pi}
    \frac{d^\frac{\alpha+\beta'}{2}}{l^\frac{\alpha+\beta'}{2}(d+l)}
\end{equation}
which applies for all $0<\alpha\leq2$ and $-1\leq\beta\leq1$.
This expression is consistent with classical results for the symmetric ($0<\alpha\leq2$, $\beta=0$) \cite[eq.\ (12)]{PhysRevLett.99.160602} and one-sided ($0<\alpha<1$, $\beta=1$) \cite[eq.\ (26)]{PhysRevLett.99.160602} \cite[eq.\ (43)]{ELIAZAR2004219} cases, and uncovers a novel one-sided case for $1<\alpha<2$, with the exponent $(\alpha+\beta')/2=\alpha-1$; that is, the first-passage leapovers on the half-line for a one-sided L\'evy process with index $1<\alpha<2$ are equivalent to those for a one-sided L\'evy process with index $\alpha-1$.

\subsection{Leapover lengths for L\'evy flights in a boundary-centred harmonic potential well}\label{sec32}

Before proceeding to the general variable transformation technique, we demonstrate it for the case of the harmonic potential well in which the restoring force is linear, which has been of particular interest in L\'evy flights \cite{PhysRevE.59.2736,Klages2008}.
We find that when the boundary lies at the extremum of the well, the FPLD (\ref{eq:fplharmonic}) can be obtained by transformation of (\ref{eq:fplkoren}).
As before we are interested in the escape from $(-\infty,d)$ for a process $X$ starting at $X_0=0$, but with stochastic differential equation
\begin{equation}
    dX = (\mu_1X+\mu_2)dt+\sigma(X,t)dL(\alpha,\beta,0,D)
\end{equation}
describing a form of L\'evy Ornstein-Uhlenbeck process \cite{PhysRevE.59.2736}.
For simplicity we assume that $\sigma(X,t)$ is positive everywhere.
Denote the first passage time $t_\mathrm{FP}:=T_{(-\infty,d)}=\inf\{t~|~X(t)\geq d\}$.
We consider the transformation defined by the function
\begin{equation}
    Y(X,t) := \left(X+\frac{\mu_2}{\mu_1}\right)e^{-\mu_1t} ~.
\end{equation}
This function has the property $dY = e^{-\mu_1t}\sigma(X,t)dL$, so $Y$ is a free L\'evy process with the time-dependent scale parameter remaining positive everywhere, with initial condition $Y(X_0,0) = \frac{\mu_2}{\mu_1}$ and boundary $Y(d,t_\mathrm{FP}) = \left(d+\frac{\mu_2}{\mu_1}\right)e^{-\mu_1t_\mathrm{FP}}$.
This boundary depends on the first passage time $t_\mathrm{FP}$, which is itself a random variable, and so the boundary is in general time-dependent, a case so far unresolved in the literature. We note that this problem is equivalent to having a constant boundary and time-dependent drift term; this can be seen by using the transformation $Z(X,t)=(X-d)e^{-\mu_1t}$ instead.

However, if $d=-\mu_2/\mu_1$, that is, the boundary of $X$ is at the centre of the harmonic potential, the boundary of $Y(X,t)$ is no longer time-dependent and resides at zero.
Given that the leapover in the original spatial variable $X$ is $l_X=X(t_\mathrm{FP})-d$, the leapover in the transformed variable $Y$ is
$l_Y=Y(X,t_\mathrm{FP})-Y(d,t_\mathrm{FP})=(X-d)e^{-\mu_1t_\mathrm{FP}}=l_Xe^{-\mu_1t_\mathrm{FP}}$.
As a result the FPLD for $Y$ is given by (\ref{eq:fplkoren}):
\begin{eqnarray}
    p_d(l_Y) dl_Y
    &= \frac{\sin(\pi\frac{\alpha+\beta'}{2})}{\pi}
    \frac{d^\frac{\alpha+\beta'}{2} dl_Y}{l_Y^\frac{\alpha+\beta'}{2}(d+l_Y)} \nonumber\\
    &= \frac{\sin(\pi\frac{\alpha+\beta'}{2})}{\pi}
    \frac{d^\frac{\alpha+\beta'}{2} e^{-\mu_1t_\mathrm{FP}}dl_X}{(e^{-\mu_1t_\mathrm{FP}}l_X)^\frac{\alpha+\beta'}{2}(d+e^{-\mu_1t_\mathrm{FP}}l_X)} ~.
\end{eqnarray}
Thus the FPLD of $X$ is
\begin{equation}\label{eq:fplharmonic}
    f_d(l,t)
    = \frac{\sin(\pi\frac{\alpha+\beta'}{2})}{\pi}
    \frac{d^\frac{\alpha+\beta'}{2}}{l^\frac{\alpha+\beta'}{2}(d+e^{-\mu_1t}l)}
    e^{-\mu_1t\left(1-\frac{\alpha+\beta'}{2}\right)}
\end{equation}
where $t$ is the first passage time.
The above measure-based transformation ensures that the FPLD remains normalised:
$\int_0^\infty f_d(l,t)dl=1$, regardless of the time at which first passage occurs.
Importantly, this FPLD is distinguished from the FPLD in the absence of a potential by the dependence on the time of first passage, despite being normalised in space for any given time, which we further explore in section \ref{sec:chechkin} and the Discussion.
The classical case of free L\'evy flights in the absence of an external potential is recovered in the limit $\mu_1\to0$.
This constitutes the first known result of FPLDs of L\'evy flights in nontrivial potential wells.

\subsection{Leapover lengths for a general class of boundary-centred potentials}

The above technique can be generalised to potentials which satisfy a certain integral condition.
Here we consider a L\'evy-driven process $X$ satisfying
\begin{equation}
    dX = \mu(X) dt + \sigma(X,t)dL(\alpha,\beta,0,D)
\end{equation}
which begins at $X_0=0$ and has boundary $d$.
Furthermore, we consider potentials $V(x)$ where the force $-V'(x)=\mu(x)$ satisfies $\int_0^X \frac{dx}{\mu(x)} = \pm\infty$ only when $X=d$, and where the function $X\mapsto\int_0^X \frac{dx}{\mu(x)}$ is (independently) injective on either side of the boundary.
As a result, there exists a constant $r$ such that
\begin{equation}\label{eq:generalfplcond}
    r\int_0^d \frac{dx}{\mu(x)} = -\infty ~.
\end{equation}
Observe that the boundary-centring condition (\ref{eq:generalfplcond}) is a stronger version of the stationary point condition $-V'(d)=\mu(d)=0$ of the potential function: the first implies the second, but the reverse is not necessarily true.
Let us consider the transformation $Y(X,t) = b(X)e^{-rt}$ for some function $b$ to be determined.
The transformed starting position is time-independent: $Y(X_0,0)=b(0)$.
The corresponding stochastic differential equation for the transformed variable is
\begin{equation}
    dY = e^{-rt}(b'(X)\mu(X)-rb(X))dt+\sigma(X,t)b'(X)e^{-rt}dL ~.
\end{equation}
To remove the drift term we impose the condition $b'(X)\mu(X)-rb(X)=0$, yielding
\begin{equation}
    |b(X)| = |b(0)|e^{r\int_0^X \frac{dx}{\mu(x)}} ~.
\end{equation}
The transformed boundary is at $Y(d,t_\mathrm{FP})=b(d)e^{-rt_\mathrm{FP}}$, so to ensure the boundary is zero (and thus non-moving) we require $b(d)=0$, which is satisfied by the boundary-centring condition (\ref{eq:generalfplcond}).

To guarantee that the boundary is properly retained in the transformed variable $Y$, the function $b$ needs to have differing sign on either side of the boundary point.
This is possible if and only if (\ref{eq:generalfplcond}) is satisfied for only the boundary point $X=d$, for if this equation were satisfied for a second point $X=d'\neq d$, then $b(d)=b(d')=0$ and so $Y(d,t)=Y(d',t)=0$, making the boundary indistinguishable from $d'$ in the transformed space.
The injectivity condition ensures that each point in the original space is uniquely represented in the transformed space, eliminating the possibility of artificial unwanted `teleports' between points in nonsingleton preimages of $Y$ with respect to $X$.

To obtain the FPLD, we note that the leapover distance for $Y$ is 
$l_Y=|Y(X,t_\mathrm{FP})|=|b(0)|e^{r(\int_0^{d+l_X}\frac{dx}{\mu(x)}-t_\mathrm{FP})}$
and so
$dl_Y=\frac{l_Yr}{\mu(d+l_X)}dl_X$.
Hence from (\ref{eq:fplkoren})
\begin{eqnarray}
    p_{|b(0)|}(l_Y)dl_Y
    &= \frac{\sin(\pi\frac{\alpha+\beta'}{2})}{\pi}
    \frac{|b(0)|^\frac{\alpha+\beta'}{2}dl_Y}{l_Y^\frac{\alpha+\beta'}{2}(|b(0)|+l_Y)} \nonumber\\
    &= \frac{\sin(\pi\frac{\alpha+\beta'}{2})}{\pi}
    \frac{|b(0)|^\frac{\alpha+\beta'}{2}}{(|b(0)|+l_Y)}
    \frac{r}{\mu(d+l_X)}l_Y^{1-\frac{\alpha+\beta'}{2}}dl_X \nonumber\\
    &= \frac{\sin(\pi\frac{\alpha+\beta'}{2})}{\pi}
    \frac{r}{\mu(d+l_X)}\frac{e^{r(\int_0^{d+l_X}\frac{dx}{\mu(x)}-t_\mathrm{FP})(1-\frac{\alpha+\beta'}{2})}}{1+e^{r(\int_0^{d+l_X}\frac{dx}{\mu(x)}-t_\mathrm{FP})}}dl_X ~.
\end{eqnarray}
Therefore the FPLD of the original process $X$ is
\begin{equation}\label{eq:fplgeneral}
    f_d(l,t)
    = \frac{\sin(\pi\frac{\alpha+\beta'}{2})}{\pi}
    \frac{r}{\mu(d+l)}\frac{e^{r(\int_0^{d+l}\frac{dx}{\mu(x)}-t)(1-\frac{\alpha+\beta'}{2})}}{1+e^{r(\int_0^{d+l}\frac{dx}{\mu(x)}-t)}} ~.
\end{equation}
In all potentials from this class the dependence on the first passage time is exponential, with growth or decay of the exponential terms depending on the shape of the potential, via the sign of $r$.
The dependence on the first passage leapover is more ambiguous, and depends on the exact shape of the potential.
The boundary-centred harmonic case is recovered when setting $\mu(x)=\mu_1x+\mu_2$.
In \ref{sec:FPLD_BC} we substitute the FPLD formula (\ref{eq:fplgeneral}) and condition on $r$ (\ref{eq:generalfplcond}) to obtain relations for $r$, $d$, and the FPLD for a number of example potentials from this class.

\section{Fractional Fokker-Planck equation with absorbing boundaries}\label{sec:chechkin}

In this section we outline the general framework for constructing the absorbing boundary condition consistently and efficiently in the FFPE, without modifying the dynamical operators of the equation.
Consider the continuity equation 
\begin{equation}\label{eq:ffpe}
    \frac{\partial P(x,t)}{\partial t} = \mathcal{A}_xP(x,t) + g(x,t)
\end{equation}
where $\mathcal{A}$, which we call the \emph{dynamical operator}, is the adjoint of the infinitesimal generator of the (Markovian) stochastic process driving the particle with position density $P(x,t)$.
The key to our framework arises from a reinterpretation of the role of the term $g(x,t)$.
This term represents an explicit effect on how the probability density function changes at a specific point in time, and is commonly referred to as a source (sink) if its magnitude is positive (negative).
For this reason, we refer to the term $g(x,t)$ as the combined \emph{source-sink}.
Through the source-sink, one can encode the behaviour of a wide variety of physical effects on particles, including boundary conditions.

The most basic type of source-sinks encountered in such equations have the spatial form of a delta function, representing the injection, or removal, of the probability density at a single point, happening at a rate determined by its coefficient, which may be time-dependent.
Another commonly seen example is a spatially extended (e.g.~Gaussian) source representing the injection of a particle at a randomly selected point determined by the form of the source function, at a rate determined by the coefficient.
Applying this principle in reverse, one finds that the effect of a spatially extended sink term is to randomly remove the particle with the relative probability of removal at a given point determined by the spatial form of the sink term, at a rate determined by its (possibly time-varying) coefficient.
As a result, the effect of a sink with the spatial form of the first-hit distribution out of a region $\Omega$, along with the FPTD as its coefficient, is to remove the particle as soon as it first exits the region $\Omega$, consistently phrasing the first passage problem for Markov processes.
This is the central result of this study: the absorbing boundary condition is consistently formulated by the usage of the first-hit distribution multiplied by the FPTD as the source-sink term, without modifying the dynamical operator $\mathcal{A}$.
This allows the usage of standard methods to analytically solve these dynamical equations, which in the case of L\'evy flights involves moving into Fourier-Laplace space, where the operator $\mathcal{A}$ has a simple formulation.
This is not possible when truncating the integral representation of the operator $\mathcal{A}$, as was classically required for the consistent phrasing of the absorbing boundary condition \cite{0305-4470-36-41-L01}.

This section is organised as follows.
In subsection \ref{sec41} we solve the FFPE (\ref{eq:ffpe}) for general Markov processes and obtain the position density (\ref{eq:markovpospdfspace}) given only the source-sink term and the free propagator.
When the dynamical operator $\mathcal{A}$ is a Fourier multiplier, the position density (\ref{eq:multiplierpospdf}) can be expressed directly in Fourier-Laplace space with knowledge of only the source-sink, and when the source-sink term is separable (e.g.\ when the first-hit distribution is time-independent) the temporal distribution of the source-sink (\ref{eq:separable_F*TD}) (e.g.\ the first passage or first arrival time densities) can be expressed in Laplace space, along with the position density (\ref{eq:separablepospdf}), using only the propagator and the spatial source-sink distribution (e.g.\ the corresponding first-hit distribution).
In subsection \ref{sec42} we consider the absorbing boundary condition for a free L\'evy flight with arbitrary skew $\beta$ out of an arbitrary region $\Omega$, and obtain the exact FPTD (\ref{eq:freelevyFPTD}) and position density (\ref{eq:freelevypospdf}) in Laplace space using only the first-hit distribution and the propagator.
Using this result, the classical test scenario of free symmetric L\'evy flight out of the half-line \cite{PhysRevE.51.2805,0305-4470-36-41-L01,PhysRevLett.99.160602,PhysRevE.86.011101} is revisited: the Sparre-Andersen scaling of the FPTD is demonstrated using elementary asymptotic considerations of the exact Laplace expression, and the position density is analytically plotted for the first time.
Returning to general Markov processes, in subsection \ref{sec43} a canonical absorbing sink term (\ref{eq:canonicalsink}) is constructed for arbitrary sources extended in both space and time, which respects the absorbing boundary condition, and modifies the source-sink in the FFPE (\ref{eq:canonicalsinkffpe}) while leaving the dynamical operator $\mathcal{A}$ unchanged.
In subsection \ref{sec44} an equation for the MFPT (\ref{eq:MFPT}) is obtained for general Markov processes using only the propagator and the first-hit distribution, even when the first-hit density is time-dependent.
This technique is used to analytically determine for the first time the MFPT (\ref{eq:MFPTfreelevy}) of free L\'evy processes of arbitrary skew $\beta$ out of a finite interval.
An implicit equation for the MFPT is also constructed in the case of L\'evy flights in the boundary-centred harmonic potential wells considered in section \ref{sec:koren}.
Finally, a discussion of how this framework might be used to analytically construct reflecting boundary conditions in the FFPE is presented in subsection \ref{sec45}.

\subsection{Framework for general Markov processes with dynamical operator $\mathcal{A}$}\label{sec41}
We begin from (\ref{eq:ffpe}), the equation describing the evolution of a Markov process with dynamical operator $\mathcal{A}$ and source-sink $g(x,t)$.
Its Laplace transform is
\begin{equation}\label{eq:ffpeLT}
    (s-\mathcal{A}_x)P(x,s) = P(x,t=0^-)+g(x,s)
\end{equation}
where $P(x,t=0^-)$ is the initial condition, assumed to be known \emph{a priori}, yielding
\begin{equation}\label{eq:markovpospdf}
    P(x,s) = G_{s-\mathcal{A}_x}(x) *_x (P(x,t=0^-)+g(x,s))
\end{equation}
where $G_{s-\mathcal{A}_x}(x)$ is the Green's function of the operator $s-\mathcal{A}_x$ (note $\mathcal{L}G_\mathcal{L}(x)=\delta(x)$), and $f(x)*_xg(x)$ denotes the convolution $\int f(x-y)g(y)dy$.
The Green's function can be reinterpreted as the free propagator $W(x,s)$, so that the position density in real space is expressed as
\begin{equation}\label{eq:markovpospdfspace}
    P(x,t) = W(x,t) *_x P(x,t=0^-)+ W(x,t)*_{x,t}g(x,t)
\end{equation}
where $*_{x,t}$ denotes combined space-time convolution.
For clarity of expression, the expressions in this section are presented in the case where the dynamical operator $\mathcal{A}_x$ is translation invariant.
When it is not, the expressions continue to apply with the following modifications: replace all instances of the Green's function $G_{s-\mathcal{A}_x}(x)$ with $G_{s-\mathcal{A}_x}(x,y)$ (where $\mathcal{L}_xG_{\mathcal{L}_x}(x,y)=\delta(x-y)$); redefine the spatial convolution to be the noncommutative operator $G(x)*_xf(x):=\int G(x,y)f(y)dy$; and replace instances of the propagator $W(x_{(b)}-x_0,\cdots)$ with $W(x_{(b)},x_0,\cdots)$.

When the source-sink $g(x,t)$ is not fully known \emph{a priori}, conditions on the position density $P(x,t)$ (such as boundary conditions) impose restrictions on the source-sink term, allowing the computation of important quantities.
For example, if there exists a point $x_b$ such that $P(x_b,t)$ is known for all time $t$, then using (\ref{eq:markovpospdf})
\begin{equation}\label{eq:BC}
    W(x_b,s)*_{x_b}P(x_b,t=0^-) = P(x_b,s)-W(x_b,s)*_{x_b}g(x_b,s)
\end{equation}
so that
\begin{equation}
    W(x_b,t)*_{x_b}P(x_b,t=0^-) = P(x_b,t)-W(x_b,t)*_{{x_b},t}g(x_b,t)
\end{equation}
which is an integral condition on the source-sink $g(x,t)$.

\subsubsection{Dynamical operator as Fourier multiplier.}
When the dynamical operator $\mathcal{A}$ can be expressed as a Fourier multiplier $\widehat{\mathcal{A}}$ (such as in free L\'evy flights), the position density can be expressed directly in Fourier-Laplace space.
Note that all multiplier operators are translation invariant.
Taking the Fourier transform of (\ref{eq:ffpeLT}) and rearranging gives
\begin{equation}\label{eq:multiplierpospdf}
    P(k,s) = \frac{P(k,t=0^-)+g(k,s)}{s-\widehat{\mathcal{A}_x}}
\end{equation}
where the free propagator $W(x,t)$ is the position density in the case of no source-sink and initial condition at zero,
\begin{equation}\label{eq:propagator}
    W(k,s) = \frac{1}{s-\widehat{\mathcal{A}_x}} ~.
\end{equation}
This concise expression (\ref{eq:multiplierpospdf}) of the position density $P(k,s)$ in Fourier-Laplace space is not obtainable using the classical formulation of the boundary conditions as a truncated integral for the dynamical operator, as in \cite{0305-4470-36-41-L01}, because those truncated integrodifferential operators are no longer Fourier multipliers.

\subsubsection{Source-sink term separable.}
Any source-sink $g(x,t)$ can be decomposed into a purely temporal part $p(t):=\int g(x,t) dx$ and a normalised spatial part $q(x,t):=g(x,t)/p(t)$ such that $g(x,t)=p(t)q(x,t)$ and $\int q(x,t)dx=1$ for all times $t\geq0$.

In many instances, such as free L\'evy flights, the source-sink $g(x,t)=p(t)q(x)$ is separable, where $q(x)$ represents the spatial distribution of injection, or removal, of probability density over time with temporal distribution $p(t)$.
For example, when $q(x)$ is the first-hit distribution out of an arbitrary region $\Omega$, and $P(x_b,t)=0$ for all $t>0$ and $x_b\notin\Omega$, then $-p(t)$ becomes the FPTD out of $\Omega$. As shown in (chechkin), when $q(x)$ is the delta sink instead then $-p(t)$ becomes the first arrival time density, in the classical sense of arrival to a point (e.g.\ the limit of zero half-width $L\to0$ in section \ref{sec:dybiec} for free L\'evy flights).

The equation for the boundary condition (\ref{eq:BC}) is rendered in the separable source-sink case as
\begin{equation}
    W(x_b,s)*_{x_b}P(x_b,t=0^-) = P(x_b,s)-W(x_b,s)*_{x_b}q(x_b)p(s)
\end{equation}
which, by rearranging, fully determines the temporal distribution of the source-sink in Laplace space
\begin{equation}\label{eq:separable_F*TD}
    p(s) = \frac{P(x_b,s)-W(x_b,s)*_{x_b}P(x_b,t=0^-)}{W(x_b,s)*_{x_b}q(x_b)}
\end{equation}
regardless whether the distribution represents first passage, first arrival, or a more exotic case.
Substituting this distribution into (\ref{eq:markovpospdf}) yields the exact expression of the position density
\begin{equation}\label{eq:separablepospdf}\fl\quad
    P(x,s) = W(x,s) *_x \left(P(x,t=0^-)
    + q(x) \frac{P(x_b,s)-W(x_b,s)*_{x_b}P(x_b,t=0^-)}{W(x_b,s)*_{x_b}q(x_b)}\right)
\end{equation}
using only the first-hit distribution $q(x)$ and propagator $W(x,s)$.

\subsection{Free L\'evy flight with absorbing BC}\label{sec42}
We now apply the above general results to fully resolve the absorbing boundary condition problem for free L\'evy flights undergoing first passage out of an arbitrary region $\Omega$, given the first-hit distribution $q_\Omega(x)$ out of $\Omega$.
Suppose that the particle starts at $x_0\in\Omega$, so $P(x,t=0^-)=\delta(x-x_0)$.
Here, the dynamical operator is a Fourier multiplier
$\widehat{\mathcal{A}}=-D|k|^\alpha(1-i\beta\tan\frac{\pi\alpha}{2}\mathrm{sgn}(k))$,
and the source-sink is separable (since the first-hit distribution is seen to be independent from time in section \ref{sec:dybiec}).
Since the boundary at $x_b\in\partial\Omega$ is absorbing, $P(x_b,t)=0$ for all $t\geq0$.

From (\ref{eq:separable_F*TD}) we obtain the exact FPTD $p_\mathrm{FP}(t)=-p(t)$ in Laplace space:
\begin{equation}\label{eq:freelevyFPTD}
    p_\mathrm{FP}(s) = \frac{W(x_b-x_0,s)}{W(x_b,s)*_{x_b}q_\Omega(x_b)}
    = \frac{F_k^{-1}\left[\frac{1}{s-\widehat{\mathcal{A}_x}}\right](x_b-x_0)}{F_k^{-1}\left[\frac{q_\Omega(k)}{s-\widehat{\mathcal{A}_x}}\right](x_b)}
\end{equation}
where $F^{-1}$ denotes the inverse Fourier transform, and the free propagator is given by (\ref{eq:propagator}):
\begin{equation}\label{eq:levypropagator}
    W(k,s) = \frac{1}{s-\widehat{\mathcal{A}_x}}
    = \frac{1}{s+D|k|^\alpha(1-i\beta\tan\frac{\pi\alpha}{2}\mathrm{sgn}(k))} ~.
\end{equation}
Using (\ref{eq:separablepospdf}) the position density in Laplace space is given by
\begin{equation}\label{eq:freelevypospdf}
    P(x,s) = W(x-x_0,s) - W(x_b-x_0,s)\frac{W(x,s)*_x q_\Omega(x)}{W(x_b,s)*_{x_b}q_\Omega(x_b)}
\end{equation}
or alternatively in Fourier-Laplace space, due to (\ref{eq:multiplierpospdf})
\begin{equation}
    P(k,s) = \frac{e^{ikx_0}-q_\Omega(k)p_\mathrm{FP}(s)}{s-\widehat{\mathcal{A}_x}} ~.
\end{equation}

This resolves the problem of first passage for free L\'evy fights of arbitrary skew parameter $\beta$ in arbitrary regions $\Omega$.
We note that the classical problem of first arrival to a point $x_b$ is recovered by setting the region $\Omega=\mathbb{R}\setminus\{x_b\}$, whereby the equations (\ref{eq:freelevyFPTD}, \ref{eq:freelevypospdf}) reduce to the well-known forms for first arrival \cite{0305-4470-36-41-L01,Palyulin2931}, thus unifying the notions of first passage and arrival.

\subsubsection{Example: free symmetric L\'evy flight, $1<\alpha<2$, $\beta=0$ \cite{PhysRevE.51.2805,0305-4470-36-41-L01,PhysRevLett.99.160602,PhysRevE.86.011101}.}
We now use the Section 2 results on the first-hit distribution to investigate the escape from the positive half-line of a particle undergoing free symmetric L\'evy flight ($x_b=0$).
The dynamical operator has a particularly simple form in this case,
$\widehat{\mathcal{A}_x} = -D|k|^\alpha$,
and the first-hit distribution is, up to a reflection, equivalent to the first passage leapover density (\ref{eq:fpl_dybiec}).
Using elementary asymptotic considerations on the FPTD (\ref{eq:freelevyFPTD}), we demonstrate in \ref{sec:sparreandersen} that the asymptotic long-time FPTD displays the Sparre-Andersen scaling
$p_\mathrm{FP}(t) \sim C(\alpha) \frac{x_0^{\alpha/2}}{\sqrt{D}}t^{-3/2}$,
where $C(\alpha)$ is a constant depending only on $\alpha$.
This result reproduces the scaling on the start-threshold distance $x_0$ and the scale parameter (generalised diffusion coefficient) $D$ previously found using a theorem due to Skorohod \cite{PhysRevLett.99.160602,Skorokhod2004}.

\begin{figure}
    \centering
    \includegraphics[width=0.47\linewidth]{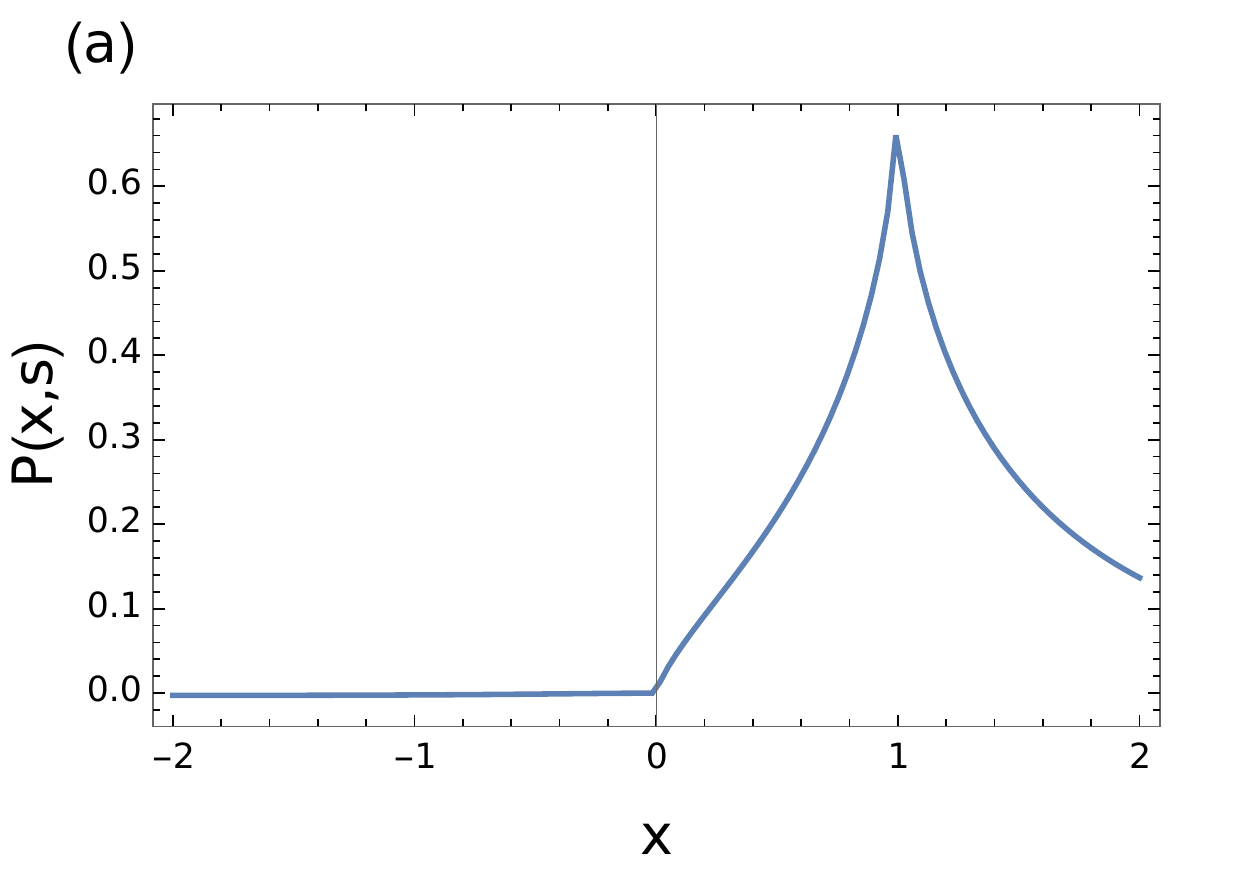}
    \includegraphics[width=0.49\linewidth]{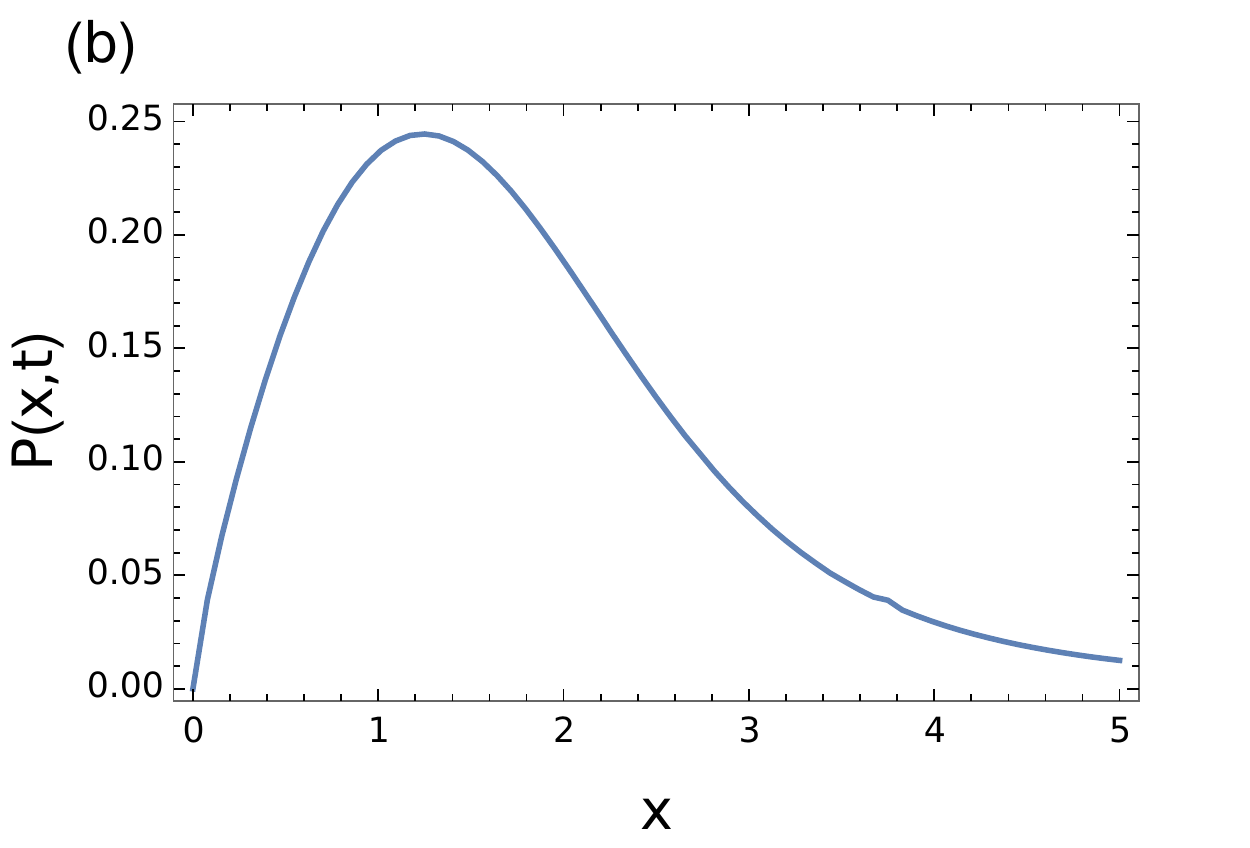}
    \caption{The analytical position density of free symmetric L\'evy flight on the positive half-line with $(\alpha,\beta,\gamma,D)=(3/2,0,0,1)$ starting at $x_0=1$. (a) The position density $P(x,s)$ in Laplace space (\ref{eq:freelevypospdf}) for $s=1$; the zero density on the negative half-line demonstrates that the absorbing boundary condition has correctly been set up. (b) The inverse Laplace transform of (\ref{eq:freelevypospdf}) yields the position density $P(x,t)$ in real space, plotted here for $t=1$. Compare with \cite[fig.\ 3]{PhysRevE.51.2805}, \cite[fig.\ 3]{PhysRevE.86.011101}.}
    \label{fig:pos}
\end{figure}

We also evaluate (\ref{eq:freelevypospdf}) in order to plot the position density of the particle in real and Laplace space (figure \ref{fig:pos}), using an implementation of the inverse Laplace transform algorithm \cite{Mathematica,horvathnumerical}.
Previously, due to the computational difficulty of solving the FFPE using the classical formulation of the absorbing boundary condition as a truncated dynamical operator, the position density had hitherto only been obtained using numerical simulations of L\'evy flights.
This efficiency problem was noted in \cite{0305-4470-36-41-L01} when establishing the truncated dynamical operator as a consistent phrasing of the FPTD problem for L\'evy flights.
Here, in order to analytically obtain the position density plots, we require only the first-hit distribution
\begin{equation}
    q_{(0,\infty)}(x) = \frac{\sin(\pi\alpha/2)}{\pi}
    \frac{|x_0|^{\alpha/2}}{|x|^{\alpha/2}|x-x_0|}
\end{equation}
supported on $x\leq0$, given by (\ref{eq:firstescape_q}) with $L\to\infty$ and $\beta=0$ (see also \cite[eq.\ (12)]{PhysRevLett.99.160602}); along with the propagator
\begin{equation}
    W(x,s) = F_k^{-1}[1/(s+D|k|^\alpha)](x) ~.
\end{equation}
The power of this novel framework is therefore demonstrated by the fact that this hitherto unknown position density (\ref{eq:freelevypospdf}) can be expressed and plotted using known, established quantities.

\subsection{Absorbing sink for arbitrary extended sources}\label{sec43}

Returning to general Markov processes, the above construction of a source-sink satisfying the absorbing boundary condition as the product of the FPTD and the first-hit distribution is only applicable for an FFPE describing a single particle, where the only source is of a delta form happening at time $t=0$.
Note that the initial condition $a(x):=P(x,t=0^-)$ can be reinterpreted as a source term $\delta(t)a(x)$ with zero initial condition everywhere.
However, in more general cases where the FFPE describes concentration densities, the source may be nontrivial and extended in both space and time, and it becomes important to determine what the corresponding extended sink term would be, such that the absorbing boundary condition remain respected.

Suppose firstly that we add $n$ particles at $(x_i,t_i)$ for $i\in\{1,\cdots,n\}$.
The source term has magnitude $\sum_{i=1}^n\delta(t-t_i)\delta(x-x_i)$ and the corresponding sink term has magnitude $\sum_{i=1}^n p_{x_i}(t-t_i)q_{x_i}(x,t-t_i)$, where $p_{x_i}(t)$ and $q_{x_i}(x)$ are the FPTD and first-hit distribution for a particle starting at $x_i$.
Thus, for an extended source $\nu(t)\mu(x,t)=\int dt_1 \nu(t_1)\delta(t-t_1)\int dx_1 \mu(x_1,t_1)\delta(x-x_1)$ the canonical absorbing sink is
\begin{eqnarray}\label{eq:canonicalsink}\fl\qquad
    S_{x,t}[\nu(t)\mu(x,t)](x,t)
    &:= \int dt_1\int dx_1\nu(t_1)p_{x_1}(t-t_1)\mu(x_1,t_1)q_{x_1}(x,t-t_1) \nonumber\\
    &= \int dx_1\nu(t)\mu(x_1,t)*_tp_{x_1}(t)q_{x_1}(x,t) ~.
\end{eqnarray}
Therefore an FFPE with arbitrary source respecting the absorbing boundary condition also contains its corresponding canonical absorbing sink as an additional (negative) contribution to the source-sink
\begin{eqnarray}\label{eq:canonicalsinkffpe}
    \frac{\partial P(x,t)}{\partial t} &= \mathcal{A}_xP(x,t) - p_{x_0}(t)q_{x_0}(x,t) \nonumber\\
    &+ \nu(t)\mu(x,t) - S_{x,t}[\nu(t)\mu(x,t)](x,t)
\end{eqnarray}
where $x_0$ is the starting position of the particle.
Note that the term $p_{x_0}(t)q_{x_0}(x,t)=S_{x,t}[\delta(t)\delta(x-x_0)]$ in (\ref{eq:canonicalsinkffpe}) is the canonical sink term in the case of a delta source in both space and time, and can be modified as appropriate for more exotic initial conditions.

\subsection{MFPT for general processes with FPLDs dependent on time}\label{sec44}

Next, we compute the MFPT for general Markov processes out of arbitrary regions $\Omega$ starting at $x_0\in\Omega$, even when their first-hit distributions are time-dependent.
The method here is inspired by a technique used in theoretical neuroscience for the analysis of networks of neurons \cite{Brunel2000,doi:10.1088/0954-898X_3_2_003,doi:10.1088/0954-898X_2_3_003,tuckwell_1988} in the Gaussian case.
We proceed by the addition of a resetting mechanism to the process, which in our case leads to the inclusion of its corresponding canonical absorbing sink term as in (\ref{eq:canonicalsinkffpe}),
\begin{eqnarray}
    \frac{\partial P(x,t)}{\partial t} &= \mathcal{A}_xP(x,t) - p_{x_0}(t)q_{x_0}(x,t) \nonumber\\\label{eq:NESSFFPE}
    &+ \nu(t)\delta(x-x_0) - \nu(t)*_t(p_{x_0}(t)q_{x_0}(x,t)) ~.
\end{eqnarray}
When the resetting rate $\nu(t)$ is equal to the rate at which the particle is absorbed (so that $P(x,t)$ maintains the same normalisation over time), we obtain the integral condition
$\nu(t) = p_{x_0}(t) + \nu(t)*_tp_{x_0}(t)$,
whose solution in Laplace space is given by $\nu(s) = p_{x_0}(s)/(1-p_{x_0}(s))$.
Physically, this means that the particle is placed at the starting position $x_0$ as soon as it is absorbed outside $\Omega$.

As time passes, a nonequilibrium stationary state (NESS) \cite{evans2019stochastic} is established in an expanding region around the resetting centre $x_0$.
Inside this region, the FPTD in the convolution term in (\ref{eq:NESSFFPE}) behaves as a delta function $p_{x_0}(t)\approx\delta(t-\langle t_\mathrm{FP}\rangle)$ centred at the MFPT.
The technique used in works such as \cite{doi:10.1088/0954-898X_2_3_003,tuckwell_1988} is to interpret the MFPT $\langle t_\mathrm{FP}\rangle$ as the inverse $1/\nu(t)$ of the resetting rate, yielding
\begin{eqnarray}\label{eq:NESSregionFFPE}
    \frac{\partial P(x,t)}{\partial t} &= \mathcal{A}_xP(x,t)
    + \nu(t)\delta(x-x_0) - \nu(t)q_{x_0}(x,t=1/\nu(t))
\end{eqnarray}
in the NESS region.
Once the system, after a large amount of time, has relaxed to the NESS, all the parameters of the system become time-independent, and the dominant contribution to the dynamics of the system arises from the MFPT $\langle t_\mathrm{FP}\rangle=1/\nu_0$, yielding from the NESS-FFPE (\ref{eq:NESSregionFFPE}) the relation
$0 = \mathcal{A}_xP_0(x)+\nu_0\delta(x-x_0)-\nu_0q_{x_0}(x,t=1/\nu_0)$
over the entire space.
As a result, the MFPT $\langle t_\mathrm{FP}\rangle$ for any Markov process with infinitesimal generator $\mathcal{A}$ starting from $x_0$ out of any region $\Omega$ is given by the solution to
\begin{equation} \label{eq:MFPT}\fl\qquad
    0 = \mathcal{A}_xP_0(x)+\frac{1}{\langle t_\mathrm{FP}\rangle}(\delta(x-x_0)
    -q_{x_0}(x,t=\langle t_\mathrm{FP}\rangle))
    ~,\qquad P_0(k=0)=1~,
\end{equation}
and so only knowledge of the dynamical operator $\mathcal{A}$ and first-hit distribution is required for the evaluation of the MFPT.
The stationary position density at $x$ represents the proportion of time spent by the particle in the region immediately around $x$, before being absorbed,
\begin{eqnarray}
    P_0(x) &= \nu_0 G_{-\mathcal{A}_x}(x) *_x (\delta(x-x_0)-q_{x_0}(x,t=1/\nu_0)) \\
    \label{eq:MFPTP0}&= \nu_0 (W(x-x_0,s=0) - W(x,s=0) *_x q_{x_0}(x,t=1/\nu_0)) ~.
\end{eqnarray}
The normalisation condition $P_0(k=0)=1$ implicitly determines $\nu_0$ and hence the MFPT.
When the infinitesimal generator of the process $\mathcal{A}$ is translation invariant, this leads to an expression for the MFPT using these quantities in Fourier space without the need for convolution operators,
\begin{equation}\label{eq:MFPTprop}
    \langle t_\mathrm{FP}\rangle = W(k=0,s=0) (1 - q_{x_0}(k=0,t=\langle t_\mathrm{FP}\rangle)) ~.
\end{equation}
It is necessary to take the value $k=0$ directly, instead of the limit $\lim_{k\to0}P_0(k)$, as we demonstrate in the following example.
Whenever the first-hit distribution is time-independent, (\ref{eq:MFPTP0}) and (\ref{eq:MFPTprop}) yield explicit expressions for the MFPT, dependent only on the propagator and the first-hit distribution.

\subsubsection{Example: free L\'evy process, escape from a finite interval \cite{PhysRevE.64.041108,BULDYREV2001148,PhysRevE.73.046104,PhysRevE.76.021116,PhysRevE.95.052102}.}
To verify the validity of this method, we find the MFPT of a particle undergoing free L\'evy flight from a finite interval $[0,2L]$, where the method yields an explicit expression.
The MFPT has been known for this situation in the symmetric case $\beta=0$ \cite{10.2307/1993412,Dybiec_2016,PhysRevE.95.052102}; using the above method we reproduce those results and extend them to arbitrary skew parameter $-1\leq\beta\leq1$, which are subsequently verified using numerical simulations.

To do this, we make use of the following result: when $q(x)\sim A_\pm/|x|^{1+\alpha}$ as $x\to\pm\infty$ in real space, then $1-q(k)\sim\frac{A_++A_-}{2}\frac{\pi}{\Gamma(1+\alpha)\sin(\pi\alpha/2)} |k|^\alpha$ for asymptotically small $k$ (positive and negative) in Fourier space.
This can be shown, for instance, by arguments similar to those in \cite[p.\ 10-11]{J.Klafter2011}.

Using (\ref{eq:levypropagator}), the implicit equation (\ref{eq:MFPTprop}) for the MFPT becomes explicit due to the first-hit distribution (\ref{eq:firstescape_q}) being time-independent:
\begin{equation}\label{eq:MFPTfreelevyeqn}
    \langle t_\mathrm{FP}\rangle = 
    \frac{1-q(k)}{D|k|^\alpha(1-i\beta\tan\frac{\pi\alpha}{2}\mathrm{sgn}(k))}\Bigg|_{k=0} ~.
\end{equation}
For large $x$, the first-hit distribution (\ref{eq:firstescape_q}) asymptotically behaves as
\begin{equation}
    q(x) \sim \frac{\sin(\frac{\pi}{2}(\alpha+\mathrm{sgn}(x)\beta'))}{\pi}
    |x_0|^\frac{\alpha-\beta'}{2} |2L-x_0|^\frac{\alpha+\beta'}{2} |x|^{-1-\alpha}
\end{equation}
so that for small $k$
\begin{equation}
    1-q(k) \sim \frac{\cos\frac{\pi\beta'}{2}}{\Gamma(1+\alpha)}
    |x_0|^\frac{\alpha-\beta'}{2} |2L-x_0|^\frac{\alpha+\beta'}{2} |k|^\alpha ~.
\end{equation}
Therefore the MFPT of free L\'evy flight from $[0,2L]$ is
\begin{equation}\label{eq:MFPTfreelevy}
    \langle t_\mathrm{FP}\rangle = \frac{\cos\frac{\pi\beta'}{2}}{D\Gamma(1+\alpha)}
    |x_0|^\frac{\alpha-\beta'}{2} |2L-x_0|^\frac{\alpha+\beta'}{2} ~.
\end{equation}

\begin{figure}
    \centering
    \includegraphics[width=0.5\linewidth]{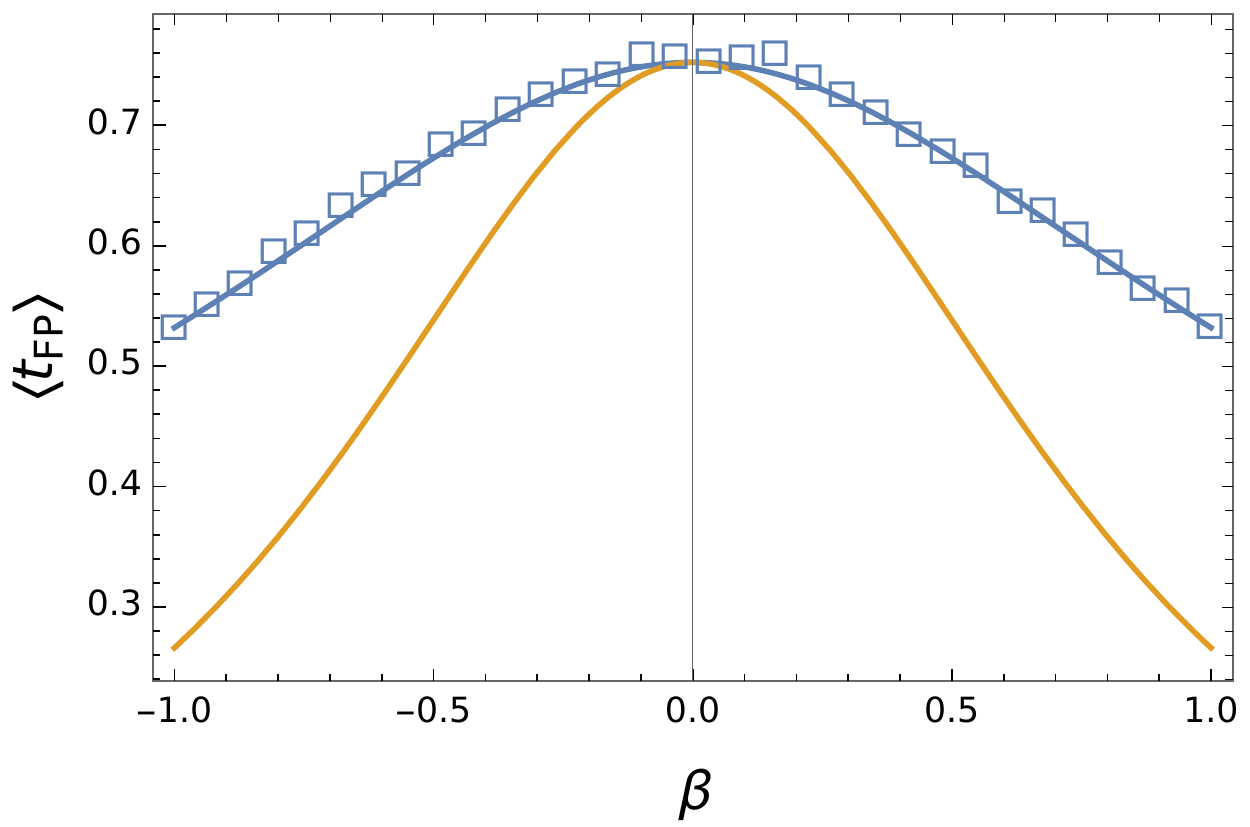}
    \caption{The MFPT (\ref{eq:MFPTfreelevy}) of free L\'evy flight (blue line) on the interval $[0,2]$ for $(\alpha,\beta,\gamma,D)=(3/2,\beta,0,1)$ starting at $x_0=L=1$, as a function of the skew parameter $\beta$. Numerical simulations (blue squares) using the Euler-Maruyama method \cite{PhysRevE.73.046104,Dybiec_2016} with $dt=10^{-3}$ averaged over $10^4$ repetitions agree with taking the value $k=0$ in the expression (\ref{eq:MFPTprop}) instead of the average of the left and right limits (yellow line).}
    \label{fig:mfpt}
\end{figure}
When $\beta=0$, the classical MFPT for the symmetric case \cite[eq.\ (1)]{Dybiec_2016} is recovered.
For the term $1-i\beta\tan\frac{\pi\alpha}{2}\mathrm{sgn}(k)$ in the denominator of (\ref{eq:MFPTfreelevyeqn}), it is necessary to take the value $k=0$ directly since the right and left limits are not equal.
Numerical simulations (figure \ref{fig:mfpt}) corroborate this finding; taking the average of the left and right limits leads to an incorrect result, which would amount to dividing (\ref{eq:MFPTfreelevy}) by $1+\beta^2\tan^2(\pi\alpha/2)$.

\subsubsection{Example: boundary-centred harmonic potential case \cite{PhysRevE.59.2736,Klages2008}.}
To demonstrate the usage of this method for more difficult cases where the first-hit distribution is time-dependent, we obtain an equation for the MFPT of a particle undergoing L\'evy flight out of the half-line in the boundary-centred harmonic potential well explored in Section \ref{sec32}.
Here, the dynamical operator $\mathcal{A}_x$ in the presence of a harmonic potential well is no longer a simple multiplier, but contains a differential operator when acting in Fourier space,
\begin{eqnarray}
    \mathcal{A}_x f(x) &= -\sigma(-\Delta)_\beta^{\alpha/2}f(x) - \frac{\partial}{\partial x} ((\mu_1x+\mu_2)f(x)) ~, \\
    \widehat{\mathcal{A}_x} &= -\sigma|k|_\beta^\alpha + \mu_1k\frac{\partial}{\partial k} + i\mu_2k ~,
\end{eqnarray}
where we assume that $\sigma$ is constant and hence, without loss of generality, $D=1$.
The operator $(-\Delta)_\beta^{\alpha/2}:=(-\Delta)^{\alpha/2}+\beta\tan\frac{\pi\alpha}{2}\frac{\partial}{\partial x}(-\Delta)^{(\alpha-1)/2} $
is the skew-adjusted fractional Laplacian with Fourier multiplier $|k|_\beta^\alpha:=|k|^\alpha(1-i\beta\tan\frac{\pi\alpha}{2}\mathrm{sgn}(k))$.
We restrict ourselves to the case $1<\alpha\leq2$; when $\alpha=1$ it suffices to replace $\tan\pi\alpha/2$ in the above Fourier multiplier expression for $|k|_\beta^\alpha$ with $-(2/\pi)\ln|k|$.

For processes such as these with nontrivial dynamical operators, it is simplest to begin with (\ref{eq:MFPT}).
We thus proceed from the equation $\mathcal{A}_xP_0(x)=h(x)$ where $h(x) := -\nu_0\delta(x-x_0)+\nu_0q(x,t=1/\nu_0)$, and the first-hit distribution $q(x,t)$ is obtained in the boundary-centred case from the FPLD (\ref{eq:fplharmonic}).
Note that the propagator $W(x,y,s=0)=P_0(x)|_{h(x)=\delta(x-y)}$, and so (\ref{eq:MFPTP0}) could be directly used instead with its noncommutative redefined spatial convolution.
Moving into Fourier space and solving the resultant ODE yields
\begin{equation}
    P_0(k) = e^{\frac{\sigma}{\alpha\mu_1}|k|_\beta^\alpha-ik\frac{\mu_2}{\mu_1}}\left(C_1+
    \int_0^ke^{\frac{-\sigma}{\alpha\mu_1}|k_1|_\beta^\alpha+ik_1\frac{\mu_2}{\mu_1}}\frac{h(k_1)}{\mu_1k_1}dk_1\right)
\end{equation}
where $C_1$ is a constant to be determined via the boundary conditions.
In real space the stationary position density
$P_0(x)=C_1P_{0,1}(x)+\frac{1}{\mu_1}P_{0,2}(x)$, where
$P_{0,1}(x)=p_{L(\alpha,\beta,0,\frac{\sigma}{\alpha\mu_1})}(x+\frac{\mu_2}{\mu_1})$ and
\begin{eqnarray}\fl\qquad
    P_{0,2}(x)
    &= \int_0^1\frac{dk_2}{k_2} p_{L(\alpha,\beta,0,\frac{\sigma}{\alpha\mu_1}(1-|k_2|^\alpha)}\left(x+\frac{\mu_2}{\mu_1}(1-k_2)\right) *_x \frac{h(x/k_2)}{k_2} \nonumber\\
    &= \int_0^1\frac{dk_2}{k_2}\int_{-\infty}^\infty dx_1 p_{L(\alpha,\beta,0,\frac{\sigma}{\alpha\mu_1}(1-|k_2|^\alpha)}\left(x+\frac{\mu_2}{\mu_1}-k_2\left(x_1+\frac{\mu_2}{\mu_1}\right)\right)h(x_1) ~.
\end{eqnarray}
In this equation, $*_x$ refers to the regular spatial convolution, while $p_{L(\alpha,\beta,\gamma,D)}$ is the probability density of the free L\'evy process with index $\alpha$, skew $\beta$, centre $\gamma$ and scale $D$.

The boundary point is at $x_b=d$ and hence $C_1=\frac{-P_{0,2}(x=d)}{\mu_1P_{0,1}(x=d)}$.
The normalisation condition $P(k=0)=1$ yields $C_1=1$ which fully and implicitly determines $\nu_0$ and hence the MFPT $1/\nu_0$.
A situation where this case is applied will be found in \cite{wardakgong}.
The normalisation condition determines the MFPT explicitly only when $\alpha=2$, since the implicit part arises from the appearance of the MFPT $1/\nu_0$ in the first-hit distribution term when $\alpha\neq2$.

\subsection{Implementing other boundary conditions}\label{sec45}

The techniques introduced in this framework can also be used to implement more exotic effects on the particle.
In this final subsection, we explore how the multiple definitions of the reflecting boundary condition \cite{PhysRevE.95.052102} may be implemented in the FFPE using extended sources and sinks without the modification of the dynamical operator.

According to \cite{PhysRevE.95.052102} there are two main types of reflecting conditions:
motion reversal or wrapping, which for wrapping-absorbing intervals is equivalent to an absorbing-absorbing interval of double the width; and
motion stopping, which is equivalent to the imposition of an infinite potential well without boundary conditions.
A pertinent question then is how these reflecting boundary conditons might be implemented in the FFPE:
\begin{enumerate}
    \item Motion reversal: Absorb the particle outside the allowed region (an absorbing sink of magnitude $p_{x_0}(t)q_{x_0}(x,t)=S_{x,t}[\delta(t)\delta(x-x_0)](x,t)$), and place it back equidistantly on the other side of the boundary (a source term of magnitude $S_{x,t}[\delta(t)\delta(x-x_0)](x_b-x,t)$).
    However, we now face the problem of managing particles which have been absorbed and replaced once and exit the region a second time, and so on. Accounting for this, we arrive at a source-sink term comprised of an infinite sum
    \begin{equation}
        \frac{\partial P}{\partial t} = \mathcal{A}P+\sum_{n=0}^\infty(-S_n(x,t)+S_n(x_b-x,t))
    \end{equation}
    where we use the recursively defined function
    \begin{eqnarray}
        S_n(x,t) &:= S_{x,t}[S_{n-1}(x_b-x,t)](x,t) ~, \\
        S_0(s,t) &:= p_{x_0}(t)q_{x_0}(x,t) = S_{x,t}[\delta(t)\delta(x-x_0)](x,t) ~,
    \end{eqnarray}
    where $S_n$ is the absorbing sink for the particle which had previously exited the region $n$ times before.
    \item Motion stopping: Absorb the particle outside the allowed region (as before) and place it on the other side of the boundary at a distance of $\epsilon$ (a source term of magnitude $p_{x_0}(t)\delta(x-(x_b+\epsilon)$).
    Managing the same problem as before leads to the infinite-sum source-sink
    \begin{equation}
        \frac{\partial P}{\partial t} = \mathcal{A}P+\sum_{n=0}^\infty\left(-S_n(x,t)+\left(\int S_n(x_1,t)dx_1\right)\delta(x-(x_b+\epsilon))\right)
    \end{equation}
    where
    \begin{eqnarray}
        S_n(x,t) &:= S_{x,t}\left[\left(\int S_{n-1}(x_1,t)dx_1\right)\delta(x-(x_b+\epsilon))\right](x,t) ~, \\
        S_0(s,t) &:= p_{x_0}(t)q_{x_0}(x,t) = S_{x,t}[\delta(t)\delta(x-x_0)](x,t) ~.
    \end{eqnarray}
\end{enumerate}
The important advantage of this approach, compared to a possible integral truncation implementation of the reflecting condition in the FFPE, is that the sinks here never depend explicitly on the position density $P(x,t)$, simplifying the solution of the FFPE.

The technique here reminds one of the scattering matrix approach used to compute scattering amplitudes in quantum field theory (pictorially represented as Feynman diagrams).

\section{Discussion}\label{sec:discussion}
% - each paper needs to be commented on somewhere
% - check papers that cited the papers relevant for this study

% ----- abstract -----
In this study we investigated the problem of Markov processes in nontrivial geometries. An analytically tractable method was devised to construct absorbing boundary conditions in their deterministic dynamical equations.
A fundamental link was established between the absorbing boundary condition and the first-hit distribution in the consideration of the first passage time problem.
Using this, an equation for the mean first passage time of Markov processes out of arbitrary regions was constructed.
For free L\'evy flights of arbitrary skew $\beta$, the first passage time density out of any region, along with the position density, was analytically expressed in Laplace space, using only the free propagator and the corresponding first-hit distribution.
With the additional importance now assigned to the first-hit distribution, the distributions of first hits and leapovers were investigated for free L\'evy flights of arbitrary skew $\beta$ in finite intervals, and L\'evy flights on the half-line in an ubiquitous class of external potentials.

% ----- section 1: introduction -----

% ----- section 2: dybiec -----
The first-hit results for free L\'evy flights in section \ref{sec:dybiec} generalise those presented recently for the symmetric case in \cite{Dybiec_2016}.
However, these escape results have been known in principle for almost half a century \cite{doi:10.1137/1117035} using coupled integral equations \cite{Profeta2016}, and those in the form presented here were obtained by a variable transformation on these classical results (see \ref{appdx:dybiec}).
This is indicative of a more general mathematician-physicist disconnect in the first passage leapover problem, where the same quantity is referred to variously as the first-hit distribution or place \cite{10.2307/1993561,doi:10.1137/1117035,lachal2007,LACHAL20081,kyprianou2014,Dybiec_2016}, the harmonic measure \cite{10.2307/1993561,Profeta2016}, and, in the case of the half-line, the FPLD \cite{ELIAZAR2004219,PhysRevLett.99.160602,KOREN200710,Dybiec_2016}.
This disconnect is evident in statements such as the claim that the prefactor of the tail of the FPTD for free symmetric L\'evy flights on the half-line was first derived in \cite{PhysRevLett.99.160602}; in fact, as mentioned in \cite{Padash_2019}, the prefactor has also been known in the mathematical literature since at least 1973 \cite{Bingham1973,10.2307/1425834} for both symmetric and asymmetric free L\'evy flights.

As in \cite{Dybiec_2016}, the two paradigmatic setups investigated in section \ref{sec:dybiec} involved the escape and arrival to finite connected intervals.
It remains to be seen whether these results are sufficient to determine the first-hit distributions for arbitrary one-dimensional regions; that is, whether these intervals can act as basis elements, in this sense, for determining the first-hit distributions for regions built out of such intervals.
Furthermore, while the original first-hit results for symmetric free L\'evy flights were presented in the multidimensional case \cite{10.2307/1993561}, the more recent results in the case of arbitrary skew restrict themselves to one-dimensional processes \cite{doi:10.1137/1117035,kyprianou2014}.
Since the first-hit distributions in the asymmetric case can be obtained using the same method as the symmetric case \cite{Profeta2016}, it should in principle be possible to present these asymmetric first-hit distribution results in the multidimensional case too.

% ----- section 3: koren -----
The FPLD for free L\'evy flights on the half-line in section \ref{sec:koren} generalise the FPLD for the symmetric case in \cite{PhysRevLett.99.160602}.
In that study, both the FPLD and the tail of the FPTD were obtained using a theorem due to Skorokhod \cite[p.\ 303]{Skorokhod2004}, which gives a formula for the joint FPTD-FPLD $p(s,u)$ in Laplace space for homogeneous processes with independent increments out of the half-line.
The FPTD and FPLD, it is said, then follow from Laplace inversion of $p(s,u=0)$ and $p(s=0,u)$ respectively \cite{PhysRevLett.99.160602}.
However, as we see from the existence of time-dependent FPLDs in section \ref{sec:koren}, the Laplace inversion of $p(s=0,u)$ actually gives the time-averaged FPLD, which is equal to the FPLD only when it is independent of time.
To obtain the FPLD when it is time-dependent, it is necessary to divide the joint distribution in real space by the FPTD, which continues to be expressed as $p(s,u=0)$ in Laplace space.
In any case, the theorem due to Skorohod is formidable in practice: the tail of the FPTD was obtained using the theorem for asymmetric L\'evy flights only recently \cite{Padash_2019}, and the FPLD for arbitrary skew (\ref{eq:fplkoren}) has not yet, to the best of our knowledge, been deduced using the theorem (cf.\ the speculative fit in \cite[fig.\ 13]{KOREN200710}).
More work needs to be done to ascertain whether the theorem can be generalised to more general classes of Markov processes, such as L\'evy flights in finite regions or those modulated by external potentials.

The first passage leapover problem for general L\'evy flights in the presence of external potential wells has not yet been explored, despite its relevance to a wide plethora of physical problems \cite{Klages2008}.
One reason for this is the sheer difficulty of the problem: the variable transformation technique in section \ref{sec:koren} reveals that the problem is equivalent to that for free L\'evy processes with time-dependent fluctuating boundaries.
The fluctuating boundary problem has not yet been fully explored for Brownian processes \cite{TaillefumierE1438,doi:10.1162/NECO_a_00577}, and the difficulty associated with the addition of only a linear drift has been noted in the case of heat-type Markov pseudoprocesses \cite{lachal2007,LACHAL20081}.
The validity of the variable transformation technique demonstrated here arises from the observation that the first passage leapover for a free L\'evy process is independent of the scale parameter, and thus these results also apply when the scale parameter is time-dependent.
The existence of time-dependent FPLDs in the case of nontrivial potential wells adds another layer of difficulty to the problem, especially when the first passage times are not known \emph{a priori}.
More generally, the variable transformation technique utilised here may be used for any stochastic process whose FPLD is scale-independent and known in the free case.

% ----- section 4: chechkin -----
The usage of the first-hit distribution as the spatially extended absorbing sink properly phrases the FPTD problem for general Markov processes in a local, pointwise manner.
The previous global formulation of the absorbing boundary condition \cite{0305-4470-36-41-L01,doi:10.1137/17M1116222}, while consistent, involved the truncation of the domain of the integral representation of the dynamical operator to the allowed region $\Omega$.
The resultant operator, in general, then lost those properties which allowed for techniques amenable to the process to be used to solve for the position density of the particle.
This issue was recognised by previous authors \cite{PhysRevE.73.046104,PhysRevE.76.021116,PhysRevE.86.011101}, and is arguably the principal reason the first passage problem for L\'evy flights remains relatively untouched compared to Brownian motion.
Physically, integral truncation corresponds to acting on the particle until it first exits the allowed region $\Omega$, at which point the particle remains stationary for all future times.
The probability density outside the region $\Omega$ then, by definition, builds up at a rate determined temporally by the FPTD, and spatially by the first-hit distribution.
Thus, subtracting this from the FFPE removes the necessity to perform integral truncation on the equation, since the inclusion of the complementary domain in the integral has no effect on the particle after the particle itself is removed.
Our formulation thus recovers the Fourier representation and translational invariance of the dynamical operators for free L\'evy flights in the presence of boundaries.
The derivation of the canonical absorbing sink term takes this into account in order to construct the absorbing boundary condition for cases where concentration densities are considered instead of discrete particles.
The first passage problem in Markov processes is thus reduced to the determination of first-hit distributions, motivating a greater focus on the problem of first hit for L\'evy flights in external potentials, and Markov processes more generally.
Whether similar techniques can be used to formulate the absorbing boundary condition problem for non-Markovian processes (e.g.\ time-fractional diffusion) remains to be elucidated.

The source-sink framework in section \ref{sec:chechkin} reconciles the notions of first passage and first arrival.
Classically, first arrival has been understood as arrival to a point, which is equivalent to first passage in the Gaussian case of Brownian motion due to its continuous sample paths.
Consequently, the method by which to distinguish between the two cases in L\'evy motion, where the discontinuous sample paths create an important distinction between first passage and arrival, has not always been clear, and has led to some confusion until very recently.
For example, while the finite-strength point sink absorbs the particle when it arrives exactly at the point, in order to construct a perfectly absorptive wall in the sense of \cite{PhysRevE.95.012154} a finite-strength Heaviside function sink (or any function with support equal to the complement of the allowed region $\Omega$) is required.
First passage, on the other hand, has classically referred to the passage out of finite or semi-infinite regions, which in both cases has an infinite region as its complement.
The approach taken in this study recognises that a particle passing out of a region arrives at its complement, and hence each of the first passage and arrival problems may be phrased in terms of the other.
For example, the classical first arrival problem can be phrased as the first passage out of the entire ambient space with the arrival point removed.
The phenomenon that the first arrival to a point is equivalent to the first passage out of the half-line in Brownian motion can then be understood in the light of the continuous sample paths of the process: when the allowed region $\Omega$ is disconnected, the Brownian motion resides in only one connected component for the entirety of the process's lifetime.
This is, of course, no longer the case in general for L\'evy processes.

The equation for the MFPT for general Markov processes is useful since the simple expression for the FPTD in Laplace space only applies when the first-hit distribution is independent of time (i.e. when the source-sink is separable).
More generally, the FPTD satisfies only an integral condition akin to (\ref{eq:BC}).
An important question that thus arises is whether this method for the MFPT can be adapted in order to compute higher-order moments of the FPTD, or the FPTD directly, possibly using techniques inspired by \cite[eqs.\ (9.135-7)]{tuckwell_1988}.
If so, the position density would then follow immediately from (\ref{eq:markovpospdfspace}) given the first-hit distribution and propagator.
As for the method of the MFPT, the application of the NESS ansatz can be understood by considering the convolution between $\nu(t)$ and $p_{x_0}(t)$ in (\ref{eq:NESSFFPE}).
In the NESS, where the resetting rate $\lim_{t\to\infty}\nu(t)>0$ is nonzero in the limit of large time, it has infinite mass compared to the finite mass of the normalised FPTD, and hence the FPTD $p_{x_0}(t)\approx\delta(x-\langle t\rangle)$ behaves as a delta function centred at the mean of the FPTD, inside the convolution.

% ----- section 5: discussion -----
The formulation of analytically efficient absorbing boundary conditions in this study has been one of the longstanding open problems of L\'evy flights.
As a result, it is hoped that this framework will contribute to the future study of L\'evy flights in nontrivial geometries, along with Markov processes more generally.
Some immediate directions for future study in terms of raw data include the usage of the exact Laplace expressions for the FPTD and position density to obtain the remaining unknown real space FPTD and position density expressions for special cases of the parameters $\alpha,\beta$ for L\'evy flights; obtaining FPLDs for more instances of boundary-centred potentials for L\'evy flights on the half-line; and deriving equations for the MFPT for more instances of Markov processes.
More generally, with the absorbing boundary condition problem for L\'evy flights now placed on a similar level of tractability with that for classical Brownian motion, the process of generalising the extensive literature on Brownian motion in bounded domains to the case of L\'evy flights can now be realistically envisioned.

\ack
The author would like to acknowledge discussions with Pulin Gong regarding a related project in theoretical neuroscience which inspired this work.
Computations were performed at the School of Physics, University of Sydney (Sydney, Australia).
This research was supported by an Australian Government Research Training Program (RTP) Scholarship.

\appendix

\section{Classical results on first arrival and passage for general $\alpha$-stable processes}\label{appdx:dybiec}

In this section, we collate a number of results on first-hit distributions for $\alpha$-stable L\'evy processes of arbitrary skew parameter $\beta$.
The variables corresponding to original formulae are primed, while the transformed setup variables used in the main section are not primed.
Some of the more recent results \cite{Profeta2016,kyprianou2014} utilise the positivity parameter $\rho=\mathbb{P}_0[X(t)>0]$ as a measure of asymmetry of the L\'evy process $X$, along with $\widehat{\rho}=1-\rho$, the positivity parameter of the dual process $\widehat{X}=-X$.
We have elected to present our results using the rescaled skewness parameter $\beta'$ defined in (\ref{eq:rescaledskew}), which provides a balance between equation readability and similarity to the skew $\beta$.
This leads to the following relations for common values of $\beta$.
When $\beta=0$, $\beta'=0$ for all $\alpha$.
The case $\beta=\pm1$ demonstrates a key difference between the two $\alpha$ regimes: when $0<\alpha<1$, $\beta'=\pm\alpha$, whereas when $1<\alpha<2$, $\beta'=\pm(\alpha-2)=\mp(2-\alpha)$.
For the sake of simplicity, the only 1-stable process considered here is the symmetric Cauchy process: $\beta=0$ when $\alpha=1$.

\cite[\textbf{theorem A(a)}]{Profeta2016}\cite[\textbf{theorem 1}]{doi:10.1137/1117035}

$|x_0'|<1$, distribution of first hits ($|x'|>1$):
\begin{equation}\label{eq:a1}\fl\qquad
    q_c(x'|x_0') = \frac{\sin(\frac{\pi}{2}(\alpha+\mbox{sgn}(x')\beta'))}{\pi}
    \left|\frac{1+x_0'}{1+x'}\right|^\frac{\alpha-\beta'}{2}
    \left|\frac{1-x_0'}{1-x'}\right|^\frac{\alpha+\beta'}{2}
    \frac{1}{|x'-x_0'|} ~.
\end{equation}
To get the first hit density on $(-\infty,0]\cup[2L,\infty)$ ($x_0\in[0,2L]$), make the transformation $x_{(0)}'\to x_{(0)}/L-1$ (where $x_{(0)}\in\{x,x_0\}$) and multiply by the Jacobian $1/L$.

\cite[\textbf{theorem A(b)}]{Profeta2016}\cite[\textbf{theorem 1}]{kyprianou2014}

$|x_0'|>1$, pdf of first hits ($|x'|<1$):
\begin{eqnarray}\label{eq:a2}\fl\qquad
    q_c(x'|x_0')
    &= \frac{\sin\left(\frac{\pi}{2}(\alpha-\mbox{sgn}(x_0')\beta')\right)}{\pi(1+x')^\frac{\alpha+\beta'}{2}(1-x')^\frac{\alpha-\beta'}{2}} \Bigg(
    \frac{|x_0'+1|^\frac{\alpha+\beta'}{2}|x_0'-1|^\frac{\alpha-\beta'}{2}}{|x'-x_0'|}
    \nonumber\\
    &\quad- \max(\alpha-1,0)\int_1^{|x_0'|}(t-1)^{\frac{\alpha-\mathrm{sgn}(x_0')\beta'}{2}-1}
    (t+1)^{\frac{\alpha+\mathrm{sgn}(x_0')\beta'}{2}-1} dt
    \Bigg) ~.
\end{eqnarray}
To get the first hit density on $[-2L,0]$ ($x_0\notin[-2L,0]$), make the transformation $x_{(0)}'\to x_{(0)}/L+1$ (where $x_{(0)}\in\{x,x_0\}$) and multiply by the Jacobian $1/L$.

\cite[\textbf{corollary 1.2}]{kyprianou2014}

$\alpha<1$, $|x_0'|>1$, probability $R_c(x_0')$ of not hitting $[-1,1]$ is
\begin{equation}\label{eq:a3}
    R_c(x_0') = \frac{\Gamma\left(1-\frac{\alpha+\mathrm{sgn}(x_0')\beta'}{2}\right)}{\Gamma\left(\frac{\alpha-\mathrm{sgn}(x_0')\beta'}{2}\right)\Gamma(1-\alpha)}
    \int_0^\frac{|x_0'|-1}{|x_0'|+1}\frac{t^{\frac{\alpha-\mathrm{sgn}(x_0')\beta'}{2}-1}}{(1-t)^\alpha}dt ~.
\end{equation}
To get the probability of not hitting $[-2L,0]$ ($x_0\notin[-2L,0]$), make the transformation $x_0\to x_0/L+1$.

Cases on the half-line for extreme $\beta$ hold by Skorokhod continuity \cite{Profeta2016}.

\section{Leapover lengths for example boundary-centred potentials}\label{sec:FPLD_BC}

Here we use (\ref{eq:fplgeneral}) to determine the FPLD for L\'evy flights in several boundary-centred potentials.
For a class of potential function, applying the boundary-centring condition (\ref{eq:generalfplcond}) yields the relation between the boundary point $d$ and the exact form of the potential function $V(x)$ with derivative $V'(x)=-\mu(x)$.

The examples presented below use formulae from the Wikipedia page ``List of integrals of rational functions''.
The injectivity condition on the integral discussed in Section \ref{sec:koren} is satisfied by functions with terms of forms such as $\log|x|$ and $1/x^n$ with $n$ even.

\begin{enumerate}
\item Regularised logarithmic potential:

Set $\mu(x)=a_1+\frac{a_2}{x}$ where $a_1\neq0$, so the L\'evy Bessel process is not included.
This corresponds to a potential well $V(x)=-(a_1x+a_2\ln|x|)$.
We find that $d=-a_2/a_1>0$ (where the potential is flat), $r=-a_1^2/a_2$, and the FPLD is
% \begin{equation}
%     f_{Y_0,\theta}(l,t_\mathrm{FP})
%     = \frac{\sin(\frac{\pi}{2}(\alpha+\beta'))}{\pi}
%     \frac{\theta+l}{\theta}
%     \frac{d^\frac{\alpha+\beta'}{2}}{l^\frac{\alpha+\beta'}{2}(d+e^{1+\frac{l-Y_0-a_1t_\mathrm{FP}}{\theta}}l)}
%     e^{(1+\frac{l-Y_0-a_1t_\mathrm{FP}}{\theta})(1-\frac{\alpha+\beta'}{2})}
% \end{equation}
\begin{equation}\fl\quad
    f_d(l,t)
    = \frac{\sin(\pi\frac{\alpha+\beta'}{2})}{\pi}
    \frac{(a_1l)^2}{d(d+l)}
    \frac{d^\frac{\alpha+\beta'}{2}}{l^\frac{\alpha+\beta'}{2}(d+e^{1+\frac{l-a_1t}{d}}l)}
    e^{(1+\frac{l-a_1t}{d})(1-\frac{\alpha+\beta'}{2})} ~.
\end{equation}
This cannot be applied to the Bessel case as when $a_1\to0$, $d\to\infty$.

\item Even-degree polynomial potential:

Suppose we have the potential well $V(x)=(a_1x+a_2)^n$, where $n\geq4$ is even.
Then $\mu(x)=-V'(x)=-na_1(a_1x+a_2)^{n-1}$, $d=-a_2/a_1$ (the centre of the potential well where it is flat), $r=-(2-n)na_1^2$, and the FPLD is
\begin{equation}\fl\quad
    f_d(l,t)
    = \frac{\sin(\pi\frac{\alpha+\beta'}{2})}{\pi}
    \frac{(2-n)na_1^2}{na_1(la_1)^{n-1}}
    \frac{e^{((a_1l)^{2-n}-(a_1d)^{2-n}+(2-n)na_1^2t)(1-\frac{\alpha+\beta'}{2})}}{1+e^{(a_1l)^{2-n}-(a_1d)^{2-n}+(2-n)na_1^2t}} ~.
\end{equation}

\item Forces of the form $f(x)/f'(x)$:

Suppose we have a potential well $V(x)$ such that the force $-V'(x)=\mu(x)=\frac{f(x)}{f'(x)}$, and suppose that $f$ has only one zero.
We find that the boundary is at that zero, i.e. $f(d)=0$, $r=1$, and
\begin{equation}\fl\quad
    f_d(l,t)
    = \frac{\sin(\pi\frac{\alpha+\beta'}{2})}{\pi}
    \frac{|f(Y_0)|^\frac{\alpha+\beta'}{2}}{|f(d+l)e^{-t}|^\frac{\alpha+\beta'}{2}(|f(Y_0)|+|f(d+l)|e^{-t})}
    |f'(d+l)|e^{-t} ~.
\end{equation}

\end{enumerate}

\section{Derivation of the FPTD for a symmetric L\'evy process on the half-line using the FFPE}\label{sec:sparreandersen}

Here we perform some rough asymptotic manipulations on the FPTD (\ref{eq:freelevyFPTD}) in Laplace space for a symmetric L\'evy process on the half-line, in order to demonstrate its Sparre-Andersen scaling on time, along with the scaling on the other variables.
We also comment on the discrepancy between the prefactor here and those obtained using previous methods.

Starting from (\ref{eq:freelevyFPTD}), we have
\begin{equation}\fl\qquad
    p_\mathrm{FP}(s) = \frac{W(-x_0,s)}{\int dx_1 q(x_1) W(-x_1,s)}
    = 1-\frac{\int dx_1 q(x_1) W(-x_1,s) - W(-x_0,s)}{\int dx_1 q(x_1) W(-x_1,s)}
\end{equation}
Now (de Moivre)
\begin{equation}
    W(x,s) = \frac{1}{2\pi}\int_{-\infty}^\infty\frac{e^{-ikx}dk}{s+D|k|^\alpha}
    =\frac{1}{\pi}\int_0^\infty\frac{\cos kx}{s+D|k|^\alpha}dk
\end{equation}
and using \cite[eq.\ (3.765.2), p.\ 438]{2007} and the identity $\Gamma(\alpha/2)\Gamma(1-\alpha/2)=\pi/\sin(\pi\alpha/2)$,
\begin{eqnarray} \fl\qquad
    \int dx_1 q(x_1) \cos(kx_1)
    = \frac{\sin\frac{\pi\alpha}{2}}{\pi}x_0^{\alpha/2}\int_{-\infty}^0\frac{dx_1\cos kx_1}{(-x_1)^{\alpha/2}(x_0-x_1)} \nonumber\\\fl\qquad
    = \frac{\sin\frac{\pi\alpha}{2}}{\pi}x_0^{\alpha/2}\int_0^\infty\frac{dx_1\cos kx_1}{x_1^{\alpha/2}(x_0+x_1)} \nonumber\\\fl\qquad
    = \frac{\sin\frac{\pi\alpha}{2}}{\pi}x_0^{\alpha/2}\frac{\Gamma(1-\alpha/2)}{2x_0^{\alpha/2}}(e^{ikx_0}\Gamma(\alpha/2,ikx_0)+e^{-ikx_0}\Gamma(\alpha/2,-ikx_0)) \nonumber\\\fl\qquad
    = \frac{1}{2\Gamma(\alpha/2)}(e^{ikx_0}\Gamma(\alpha/2,ikx_0)+e^{-ikx_0}\Gamma(\alpha/2,-ikx_0))
\end{eqnarray}
By Fubini and the relation $\Gamma(a,x)=\Gamma(a)-\gamma(a,x)$,
\begin{eqnarray}
    \int dx_1q(x_1)W(-x_1,s)
    = \frac{1}{\pi}\int_0^\infty\frac{\int dx_1q(x_1)\cos kx_1}{s+D|k|^\alpha}dk \nonumber\\
    = \frac{1}{2\Gamma(\alpha/2)\pi}\int_{-\infty}^\infty\frac{e^{ikx_0}\Gamma(\alpha/2,ikx_0)}{s+D|k|^\alpha}dk \nonumber\\
    = W(-x_0,s)-\frac{1}{2\Gamma(\alpha/2)\pi}\int_{-\infty}^\infty\frac{e^{ikx_0}\gamma(\alpha/2,ikx_0)}{s+D|k|^\alpha}dk
\end{eqnarray}
Thus
\begin{equation}\fl\qquad
    p_\mathrm{FP}(s)
    = 1 - \frac{\frac{-1}{2\Gamma(\alpha/2)\pi}\int_{-\infty}^\infty\frac{e^{ikx_0}\gamma(\alpha/2,ikx_0)}{s+D|k|^\alpha}dk}{W(-x_0,s)-\frac{1}{2\Gamma(\alpha/2)\pi}\int_{-\infty}^\infty\frac{e^{ikx_0}\gamma(\alpha/2,ikx_0)}{s+D|k|^\alpha}dk}
    =: 1-\frac{\mathrm{num.}}{\mathrm{denom.}}
\end{equation}
For the numerator, note that
\begin{equation}
    \int_{-\infty}^\infty\frac{e^{ikx_0}\gamma(\alpha/2,ikx_0)}{s+D|k|^\alpha}dk
    = \int_{-L}^L\frac{e^{ikx_0}\gamma(\alpha/2,ikx_0)}{s+D|k|^\alpha}dk + R_1(L)
\end{equation}
where $R_1(L)\to0$ as $L\to\infty$ independently of $s$ using the dominated convergence theorem.
Using the following power series expansion of the lower incomplete gamma function
\begin{equation}
    \gamma(a,z)=\Gamma(a)e^{-z}\sum_{n=0}^\infty\frac{z^{n+a}}{\Gamma(a+n+1)}
\end{equation}
we obtain
\begin{eqnarray}
    \mathrm{num.}
    = \frac{-1}{2\Gamma(\alpha/2)\pi}
    \int_{-L}^L\frac{e^{ikx_0}\Gamma(\alpha/2)e^{-ikx_0}\sum_{n=0}^\infty\frac{(ikx_0)^{n+\alpha/2}}{\Gamma(\alpha/2+n+1)}}{s+D|k|^\alpha}dk + R_1(L) \nonumber\\
    = \frac{-1}{2\pi}\sum_{n=0}^\infty\frac{(ix_0)^{n+\alpha/2}}{\Gamma(\alpha/2+n+1)}\int_{-L}^L\frac{k^{n+\alpha/2}}{s+D|k|^\alpha}dk + R_1(L)
\end{eqnarray}
Now
\begin{eqnarray}
    \int\frac{k^bdk}{s+Dk^\alpha} = \frac{k^{b+1}~_2F_1(1,\frac{b+1}{\alpha};1+\frac{b+1}{\alpha};-Dk^\alpha/s)}{s(b+1)}
    + \mathrm{const.} \nonumber\\
    \sim s^{\frac{b+1}{\alpha}-1}D^{-\frac{b+1}{\alpha}}\frac{\Gamma(1-\frac{b+1}{\alpha})\Gamma(1+\frac{b+1}{\alpha})}{b+1}\phi_b \label{eq:mathematica}
\end{eqnarray}
as $s\to0$, where $\phi_b=e^{-\frac{2\pi i(b+1)}{\alpha}\lfloor\frac{\pi+\arg(s)-\arg(Dk^\alpha)}{2\pi}\rfloor}=1$ when $k$ is nonnegative.
Thus
\begin{eqnarray}
    \int_{-L}^L\frac{k^{n+\alpha/2}}{s+D|k|^\alpha}dk
    = (1+(-1)^{n+\alpha/2})\int_0^L\frac{k^{n+\alpha/2}}{s+Dk^\alpha}dk \nonumber\\
    \sim (1+(-1)^{n+\alpha/2})s^{\frac{n+1}{\alpha}-\frac{1}{2}}D^{-\frac{n+1}{\alpha}-\frac{1}{2}}\frac{\Gamma(\frac{1}{2}-\frac{n+1}{\alpha})\Gamma(\frac{3}{2}+\frac{n+1}{\alpha})}{n+1+\alpha/2}
\end{eqnarray}
as $s\to0$, independently of $L$, and so
\begin{eqnarray}\fl\qquad
    \mathrm{num.}
    \sim \frac{-1}{2\pi}\sum_{n=0}^\infty\frac{(ix_0)^{n+\alpha/2}(1+(-1)^{n+\alpha/2})}{\Gamma(\alpha/2+n+1)}
    s^{\frac{n+1}{\alpha}-\frac{1}{2}}D^{-\frac{n+1}{\alpha}-\frac{1}{2}}\frac{\Gamma(\frac{1}{2}-\frac{n+1}{\alpha})\Gamma(\frac{3}{2}+\frac{n+1}{\alpha})}{n+1+\alpha/2} \nonumber\\\fl\qquad
    \sim \frac{-1}{2\pi}\frac{(ix_0)^{\alpha/2}}{\Gamma(\alpha/2+1)}
    (1+(-1)^{\alpha/2})s^{\frac{1}{\alpha}-\frac{1}{2}}D^{-\frac{1}{\alpha}-\frac{1}{2}}\frac{\Gamma(\frac{1}{2}-\frac{1}{\alpha})\Gamma(\frac{3}{2}+\frac{1}{\alpha})}{1+\alpha/2}
\end{eqnarray}
where the last asymptotic relation was obtained by taking the leading term in $s$.
For the denominator, note that
\begin{equation}
    W(-x_0,s)
    = \frac{1}{\pi}\int_0^\infty\frac{\cos kx_0}{s+D|k|^\alpha}dk
    = \frac{1}{\pi}\int_0^L\frac{\cos kx_0}{s+D|k|^\alpha}dk + R_2(L)
\end{equation}
where, similarly to $R_1(L)$, using the dominated convergence theorem 
$R_2(L)=\frac{1}{\pi}\int_L^\infty\frac{\cos kx_0}{s+D|k|^\alpha}dk\to\frac{1}{\pi}\int_L^\infty\frac{\cos kx_0}{D|k|^\alpha}dk\to0$
as $s\to0$ and $L\to\infty$ respectively.
Now
\begin{eqnarray}
    \frac{1}{\pi}\int_0^L\frac{\cos kx_0}{s+D|k|^\alpha}dk
    = \frac{1}{\pi}\int_0^L\frac{\sum_{n=0}^\infty\frac{(-1)^n(kx_0)^{2n}}{(2n)!}}{s+D|k|^\alpha}dk \nonumber\\
    = \frac{1}{\pi}\sum_{n=0}^\infty\frac{(-1)^nx_0^{2n}}{(2n)!}\int_0^L\frac{k^{2n}}{s+D|k|^\alpha}dk \nonumber\\
    \sim \frac{1}{\pi}\sum_{n=0}^\infty\frac{(-1)^nx_0^{2n}}{(2n)!}
    s^{\frac{2n+1}{\alpha}-1}D^{-\frac{2n+1}{\alpha}}\frac{\Gamma(1-\frac{2n+1}{\alpha})\Gamma(1+\frac{2n+1}{\alpha})}{2n+1}
\end{eqnarray}
as $s\to0$ independently of $L$.
As a result, taking the leading term in $s$,
\begin{equation}
    W(-x_0,s)
    \sim \frac{1}{\pi}s^{\frac{1}{\alpha}-1}D^{-\frac{1}{\alpha}}\Gamma(1-\frac{1}{\alpha})\Gamma(1+\frac{1}{\alpha})
\end{equation}
and so this term dominates in the denominator as $s\to0$.
Therefore
\begin{eqnarray}
    p_\mathrm{FP}(s)
    \sim 1 + \frac{1}{2} \frac{(ix_0)^{\alpha/2}}{\Gamma(\alpha/2+1)}
    \frac{1+(-1)^{\alpha/2}}{1+\alpha/2} s^{1/2}D^{-1/2}
    \frac{\Gamma(\frac{1}{2}-\frac{1}{\alpha})\Gamma(\frac{3}{2}+\frac{1}{\alpha})}{\Gamma(1-\frac{1}{\alpha})\Gamma(1+\frac{1}{\alpha})} \nonumber\\
    = 1 + \left(\cos\frac{\pi\alpha}{4}\tan\frac{\pi}{\alpha}\right)\frac{2s^{1/2}}{\alpha\sqrt{D}\Gamma(\alpha/2)}x_0^{\alpha/2}
\end{eqnarray}
from which we obtain
\begin{eqnarray}
    p_\mathrm{FP}(t) \sim \left(-\cos\frac{\pi\alpha}{4}\tan\frac{\pi}{\alpha}\right)\frac{x_0^{\alpha/2}}{\alpha\sqrt{\pi D}\Gamma(\alpha/2)} t^{-3/2}
\end{eqnarray}
where the prefactors are expressed in a form suitable for comparison with the prefactors obtained from other methods \cite{Bingham1973,10.2307/1425834,PhysRevLett.99.160602,Padash_2019}.

Even though the derivation here is rough, with the key asymptotic result (\ref{eq:mathematica}) obtained using computer algebra software \cite{Mathematica}, it is important to note that all previous methods have ultimately derived, to the knowledge of the author, from techniques due to Lamperti, either using the theorem due to Skorokhod \cite{Skorokhod2004}, the formalised Lamperti transform \cite{kyprianou2014}, or directly \cite{Bingham1973,10.2307/1425834}; all of which make use of similar intermediate expressions in order to arrive at the prefactor.
However, the approach here used elementary techniques on the first-hit distribution along with the free propagator.
It is hoped that future work will provide a cleaner asymptotic expansion of (\ref{eq:freelevyFPTD}) in order to recover the correct prefactor.

\section*{References}
\bibliographystyle{unsrt}
\bibliography{main.bib}

%%end novalidate
\end{document}